\documentclass{llncs}

\usepackage{amsmath,amssymb,latexsym,stmaryrd}
\usepackage[algoruled,vlined,english]{algorithm2e}
\usepackage{galois}
\usepackage[pdftex]{graphicx}
\usepackage{caption}
\usepackage{bm}

\usepackage[pdftex,%
  	colorlinks,%
  	pdfauthor={Ganty, Majumdar, Monmege},%
  	hypertexnames=false,%
		plainpages=false,%
  	pdftitle={Bounded Underapproximations}]{hyperref}

\def\@envspa{\hspace{0.3em}}
\def\@sa{\hspace{-0.2em}}
\def\@sb{\hspace{0.5em}}
\def\@sc{\hspace{-0.1em}}

\makeatletter
\begingroup \catcode `|=0 \catcode `[= 1
\catcode`]=2 \catcode `\{=12 \catcode `\}=12
\catcode`\\=12 |gdef|@xcomment#1\end{comment}[|end[comment]]
|endgroup
\def\@comment{\let\do\@makeother \dospecials\catcode`\^^M=10\def\par{}}
\def\begincomment{\@comment\@xcomment}
\makeatother
\newenvironment{comment}{\begincomment}{}

\newif
  \iflong
  \longtrue
\newif
  \ifshort
	\shortfalse

\def\ie{{\it i.e. }}

\def\set#1{{\left\{ #1 \right\}}}
\def\tuple#1{{\left\langle #1 \right\rangle}}
\def\nats{{\mathbb{N}}}
\def\integers{{\mathbb{Z}}}

\def\pdec{{\sf P1}}
\def\pmany{{\sf P2}}
\newcommand{\zero}{\ensuremath{\bar{0}}}
\newcommand{\one}{\ensuremath{\bar{1}}}
\def\pzero{\boldsymbol{0}}
\def\vzero{\ddot{0}}
\def\punit{\mathbf{e}}
\def\ring{\ensuremath{\mathcal{S}}}
\def\lring{\ensuremath{\mathcal{L}}}
\def\pring{\ensuremath{\mathcal{P}}}
\def\deriv#1{#1|}
\def\sizeof#1{\mathit{sizeof}({#1})}
\def\funbf{\boldsymbol{F}}
\def\bnu{\boldsymbol{\nu}}
\def\bkappa{\boldsymbol{\kappa}}
\def\bmu{\boldsymbol{\mu}}
\def\comb{\ensuremath{\oplus}}
\def\bigcomb{\ensuremath{\bigoplus}}
\def\ext{\ensuremath{\odot}}
\def\vbf{\ensuremath{\boldsymbol{v}}}
\def\wbf{\ensuremath{\boldsymbol{w}}}
\def\post{\ensuremath{\mathrm{post}}}

\begin{document}

\title{Bounded Underapproximations}

\author{Pierre Ganty%
\texorpdfstring{\inst{1}}{1} , %
Rupak Majumdar%
\texorpdfstring{\inst{2}}{2} , %
Benjamin Monmege%
\texorpdfstring{\inst{3}}{3}}


\ifshort
\institute{$^1\;$IMDEA Software, Spain\quad  $^2\;$UCLA, USA\quad $^3\;$ENS Cachan, France}
\fi

\iflong
\institute{UCLA, USA \email{rupak@cs.ucla.edu} \and IMDEA Software, Spain \email{pierre.ganty@imdea.org}\and  ENS Cachan, France \email{bmonmege@dptinfo.ens-cachan.fr}}
\fi

\maketitle

\begin{abstract}
We show a new and constructive proof of the following language-theoretic
result: for every context-free language $L$, there is a {\em bounded} context-free language
$L'\subseteq L$ which has the same Parikh (commutative) image as $L$.  Bounded
languages, introduced by Ginsburg and Spanier, are subsets of
regular languages of the form $w_1^*w_2^*\cdots w_k^*$ for some
$w_1,\ldots,w_k\in\Sigma^*$. In particular bounded subsets of context-free
languages have nice structural and decidability
properties.  Our proof proceeds in two parts. First, using Newton's iterations
on the language semiring, we construct a context-free subset $L_N$ of $L$ that
can be represented as a sequence of substitutions on a linear language and has
the same Parikh image as $L$.  Second, we inductively construct
a Parikh-equivalent bounded context-free subset of $L_N$.

\ifshort
As an application of this result in model checking, we show how to
underapproximate the reachable state space of multithreaded procedural
programs. The bounded language constructed above provides a decidable
underapproximation for the original problem.
\fi
\iflong
We show two applications of this result in model checking: to underapproximate
the reachable state space of multithreaded procedural programs and to
underapproximate the reachable state space of recursive counter programs.  The
bounded language constructed above provides a decidable underapproximation for
the original problems.
\fi
By iterating the construction, we get a semi-algorithm
for the original problems that constructs a sequence of underapproximations
such that no two underapproximations of the sequence can be compared.  This
provides a progress guarantee: every word $w\in L$ is in some underapproximation
of the sequence, and hence, a program bug is guaranteed to be found.
In particular, we show that verification with bounded languages
generalizes context-bounded reachability for multithreaded programs.
\end{abstract}

\section{Introduction}

Many problems in program analysis reduce to undecidable problems about context-free
languages. For example, checking safety properties of multithreaded recursive
programs reduces to checking emptiness of the intersection of context-free
languages \cite{Ramalingam,BET03}. 
\iflong %
Checking reachability for recursive
counter programs relies on context-free languages to describe valid control
flow paths.
\fi 

We study underapproximations of these problems, with the intent of building
tools to find bugs in systems.
In particular, we study underapproximations in which one or
more context-free languages arising in the analysis are replaced by their
subsets in a way that
(\pdec) the resulting problem after the replacement becomes decidable
and
(\pmany) the subset preserves ``many'' strings from the original language.
Condition (\pdec) ensures that we have an algorithmic check for the
underapproximation.  Condition (\pmany) ensures that we are likely to retain
behaviors that would cause a bug in the original analysis.

We show in this paper an underapproximation scheme using {\em bounded
languages} \cite{GS,ginsburg}.  A language $L$ is {\em bounded} if
there exist $k\in\nats$ and finite words $w_1, w_2,\ldots,w_k$ such that $L$ is
a subset of the regular language $w_1^*\cdots w_k^*$.  In particular,
context-free bounded languages (hereunder bounded languages for short) have
stronger properties than general context-free languages: for example, it is
decidable to check if the intersection of a context-free language and a bounded
language is non-empty \cite{GS}.  For our application to verification, these
decidability results ensure condition (\pdec) above.

The key to condition (\pmany) is the following {\em Parikh-boundedness}
property: for every context-free language $L$, there is a bounded language
$L' \subseteq L$ such that the Parikh images of $L$ and $L'$ coincide.
(The {\em Parikh image} of a word $w$ maps each symbol of the alphabet to the number of
times it appears in $w$, the Parikh image of a language is the set of Parikh
images of all words in the language.) A language $L'$ meeting the above
conditions is called a {\em Parikh-equivalent bounded subset} of $L$.
Intuitively, $L'$ preserves ``many'' behaviors as for every string in $L$,
there is a permutation of its symbols that matches a string in $L'$.

The Parikh-boundedness property was first proved in \cite{LL79,BL81}, however,
the chain of reasoning used in these papers made it difficult to see how to
explicitly construct the Parikh-equivalent bounded subset.  Our paper gives a
direct and constructive proof of the theorem.  We identify 
\iflong
three 
\else
two
\fi
contributions in this paper.

\noindent  %
{\bf Explicit construction of Parikh-equivalent bounded subsets.} Our
constructive proof falls into two parts.  First, using Newton's iteration
\cite{EKL07} on the semiring of languages, we construct, for a given
context-free language $L$, a finite sequence of linear substitutions which
denotes a Parikh-equivalent (but not necessarily bounded) subset of $L$. (A
linear substitution maps a symbol to a language defined by a {\em linear}
grammar, that is, a context-free grammar where each rule has at most one
non-terminal on the right-hand side.) The Parikh equivalence follows from a
convergence property of Newton's iteration.

Second, we provide a direct constructive proof that takes as input such a
sequence of linear substitutions, and constructs by induction a
Parikh-equivalent bounded subset of the language denoted by the sequence.

\smallskip %
\noindent
{\bf Reachability analysis of multithreaded programs with procedures.} Using
the above construction, we obtain a semi-algorithm for reachability analysis of
multithreaded programs with the intent of finding bugs.  To check if
configuration $(c_1,c_2)$ of a recursive 2-threaded program is reachable, we
construct the context-free languages $L^0_1 = L(c_1)$ and $L^0_2 = L(c_2)$
respectively given by the execution paths whose last configurations are $c_1$
and $c_2$, and check if either $L_1' \cap L^0_2$ or $L^0_1\cap L_2'$ is
non-empty, where $L_1'=L^0_1 \cap w_1^*\cdots w_k^*$ and $L_2'=L^0_2\cap
v_1^*\cdots v_l^*$ are two Parikh-equivalent bounded subsets of $L^0_1$ and
$L^0_2$, respectively.  If either intersection is non-empty, we have found a
witness trace.  Otherwise, we construct $L^1_1 = L^0_1 \cap
\overline{w_1^*\cdots w_k^*}$ and $L^1_2 = L^0_2 \cap \overline{v_1^*\cdots
v_l^*}$ in order to exclude, from the subsequent analyses, the execution paths
we already inspected. We continue by rerunning the above analysis on $L^1_1$
and $L^1_2$.  If $(c_1,c_2)$ is reachable, the iteration is guaranteed to
terminate; if not, it could potentially run forever.  Moreover, we show our
technique subsumes and generalizes context-bounded reachability \cite{QR}.

\iflong
\noindent %
{\bf Reachability analysis of programs with counters and procedures.} We also
show how to underapproximate the set of reachable states of a procedural
program that manipulates a finite set of counters.  This program is given as a
counter automaton $A$ (see \cite{ls04} for a detailed definition) together with
a context-free language $L$ over the transitions of $A$. Our goal is to compute
the states of $A$ that are reachable using a sequence of transitions in $L$.

A possibly non terminating algorithm to compute the reachable states of $A$
through executions in $L$ is to (1) find a Parikh-equivalent bounded subset
$L'$ of $L$; (2) compute the states that are reachable using a sequence of
transitions in $L'$ (as explained in \cite{ls04}, this set is computable if
$(i)$ some restrictions on the transitions of $A$ ensures the set is Presburger
definable and $(ii)$ $L'$ is bounded, i.e.  $L'\subseteq w_1^*\cdots w_k^*$);
and (3) rerun the
analysis using for $L\cap\overline{w_1^*\cdots w_k^*}$ so that runs already
inspected are omitted in every subsequent analyses.
Again, every path in $L$ is eventually covered in the iteration.
\fi

\ifshort
\noindent
In \cite{CoRR}, we provide detailed proofs and one more applications of our result.
\fi

\noindent%
{\bf Related Work.}
Bounded languages have been recently proposed by Kahlon for tractable
reachability  analysis of multithreaded  programs \cite{Kahlon}.  His
observation is that in many practical instances of multithreaded reachability,
the languages are actually bounded.  If this is true, his algorithm checks the
emptiness of the intersection (using the algorithm in \cite{GS}).  In 
contrast, our results are applicable even if the boundedness property does not hold.

For multithreaded reachability, {\em context-bounded reachability}
\cite{QR,SES08} is a popular underapproximation technique which tackles the
undecidability by limiting the search to those runs where the active thread
changes at most $k$ times.  Our algorithm using bounded languages {\em
subsumes} context-bounded reachability, and can capture unboundedly many
synchronizations in one analysis.  We leave the empirical evaluation of our
algorithms for future work.

\section{Preliminaries}

An alphabet is a finite non-empty set of symbols. We use the letter $\Sigma$ to
denote some alphabet.  We assume the reader is familiar with the basics of
language theory (see \cite{HU79}).  The concatenation $L\cdot L'$
of two languages $L,L'\subseteq \Sigma^*$ is defined using word concatenation
as $L\cdot L'=\set{l \cdot l'\mid l\in L\land l'\in L'}$.

An {\em elementary bounded language} over $\Sigma$ is a language of the form
$w_1^*\cdots w_k^*$ for some $w_1,\ldots,w_k\in \Sigma^*$.

\noindent %
{\bf Vectors.}
For $p\in\nats$, we write $\integers^p$ and $\nats^p$ for the set of $p$-dim
vectors (or simply vectors) of integers and naturals, respectively. We write
$\pzero$ for the vector $(0,\dots,0)$ and $\punit_i$ the vector $(z_1,\dots,z_p)\in\nats^p$
such that $z_j = 1$ if $j = i$ and $z_j = 0$ otherwise.
\emph{Addition}
on $p$-dim vectors is the
componentwise extension of its scalar counterpart, that is, given
$(x_1,\dots,x_p),(y_1,\dots,y_p)\in\integers^p$
$(x_1,\dots,x_p)+(y_1,\dots,y_p)=(x_1+y_1,\dots,x_p+y_p)$.
\iflong  
Given $\lambda\in\nats$ and
$x\in\integers^p$, we write $\lambda x$ as the
$\lambda$-times sum
$x + \dots + x$.
\fi

\noindent %
{\bf Parikh Image.}
Give $\Sigma$ a fixed linear order:
$\Sigma=\set{a_1,\ldots,a_p}$.  The Parikh image of a symbol $a_i\in\Sigma$,
written $\Pi_{\Sigma}(a_i)$, is $\punit_i$.
The Parikh image is extended to words of $\Sigma^*$ as follows:
$\Pi_{\Sigma}(\varepsilon)=\pzero$ and $\Pi_{\Sigma}(u\cdot
v)=\Pi_{\Sigma}(u)+\Pi_{\Sigma}(v)$.  Finally, the Parikh image of a language
on $\Sigma^*$ is the set of Parikh images of its words.
We also define, using vector addition, the operation $\dotplus$ on
sets of Parikh vectors as follows: given $Z,Z'\subseteq\nats^p$,
let $Z\dotplus Z'=\set{z + z'\mid z\in Z\land z'\in Z'}$.
Thus, $\Pi_{\Sigma}$ maps $2^{\Sigma^*}$ to $2^{\nats^p}$.
\ifshort
When it is clear we omit
the subscript in $\Pi_{\Sigma}$.
\fi
\iflong
We also define the inverse of the Parikh image
$\Pi_{\Sigma}^{-1}\colon 2^{\nats^p}\rightarrow
2^{\Sigma^*}$ as follows: given a subset $M$ of $\nats^{p}$,
$\Pi^{-1}_{\Sigma}(M)$ is the set $\set{y\in\Sigma^*\mid \exists
m\in M\colon m=\Pi_{\Sigma}(y)}$.  When it is clear from the context we generally omit
the subscript in $\Pi_{\Sigma}$ and $\Pi^{-1}_{\Sigma}$.

The following lemma gives the properties of $\Pi$ and $\Pi^{-1}$ we need in the
sequel.
\begin{lemma}
For every $M\in 2^{\nats^{p}}$ we have $\Pi\comp\Pi^{-1}(M)=M$.\\
Let $\phi=\Pi^{-1}\comp \Pi$, for every $X,Y\subseteq\Sigma^*$ we have:
\vspace{-7pt}
\begin{description}
	\item[additivity of $\Pi$] $\Pi(X\cup Y)=\Pi(X)\cup\Pi(Y)$;
	\item[monotonicity of $\phi$] $X\subseteq Y$ implies $\phi(X)\subseteq \phi(Y)$;
	\item[extensivity of $\phi$] $X\subseteq\phi(X)$;
	\item[idempotency of $\phi$] $\phi\comp\phi(X)=\phi(X)$;
	\item[structure-semipreservation of $\phi$] $\phi(X)\cdot \phi(Y)\subseteq\phi(X\cdot Y)$;
	\item[preservation of $\Pi$] $\Pi(X\cdot Y)=\Pi(X)\dotplus\Pi(Y)$.
\end{description}
\label{lem:additiveinsertion}
\end{lemma}
\begin{proof}
	For the first statement we first observe that $\Pi$ is a
	surjective function, for each vector of $\nats^p$ there is a word that
	is mapped to that vector. Next,
	\begin{align*}
		\Pi\comp\Pi^{-1}(M)&=\Pi(\set{y\mid \exists m\in M\colon m=\Pi(y)})&\text{def.\ of $\Pi^{-1}$}\\
		&=\set{\Pi(y)\mid \exists m\in M\colon m=\Pi(y)}&\text{def.\ of $\Pi$}\\
		&= M &\text{surjectivity of $\Pi$}
	\end{align*}
	For the additivity, the monotonicity, the extensivity and the idempotency properties, we simply show the equivalence given below. Hence the properties immediately follows by property of Galois connection (we refer the reader to \cite{cousotphdthesis} for detailed proofs).
We show that for every $L\in2^{\Sigma^*}, M\in 2^{\nats^{\Sigma}}$ we have :
$\Pi(L) \subseteq M$ if{}f $L\subseteq \Pi^{-1}(M)$.
	\begin{align*}
		&L \subseteq \Pi^{-1}(M)\\
		&\text{ if{}f }L \subseteq\set{y\mid \exists m\in M\colon m=\Pi(y)} &\text{def.\ of $\Pi^{-1}$}\\
		&\text{ if{}f }\forall \ell\in L \,\exists m\in M\colon m=\Pi(\ell)\\
		&\text{ if{}f }\forall h\in \Pi(L) \,\exists m\in M\colon m=h&\text{def.\ of $\Pi$}\\
		&\text{ if{}f } \Pi(L)\subseteq M
	\end{align*}
For structure semipreservation, we prove that $\phi(x)\cdot
\phi(y)\subseteq\phi(x\cdot y)$ for $x,y\in \Sigma^*$ as follows:
	\begin{align*}
		\phi(x)\cdot\phi(y)&=\Pi^{-1}\comp\Pi(x)\cdot \Pi^{-1}\comp\Pi(y)\\
		&=\set{x'\mid \Pi(x')=\Pi(x)}\cdot\set{y'\mid \Pi(y')=\Pi(y)}&\text{def.\ of $\Pi^{-1}$}\\
		&=\set{x'\cdot y'\mid \Pi(x')=\Pi(x)\land \Pi(y')=\Pi(y)}\\
		&\subseteq\set{x'\cdot y'\mid \Pi(x')+\Pi(y')=\Pi(x)+\Pi(y)}\\
		&=\set{x'\cdot y'\mid \Pi(x'\cdot y')=\Pi(x\cdot y)}&\text{def.\ of $\Pi$}\\
		&=\Pi^{-1}\comp\Pi(x\cdot y)\\
		&=\phi(x\cdot y)
	\end{align*}
	The result generalizes to languages in a natural way. Finally, the preservation
	of $\Pi$ is proved as follows:
	\begin{align*}
		\Pi(X\cdot Y)&=\set{\Pi(w)\mid w\in X\cdot Y} &\text{def.\ of $\Pi$}\\
		&=\set{\Pi(x\cdot y)\mid x\in X\land y\in Y} &\text{def.\ of $\cdot$}\\
		&=\set{\Pi(x)+\Pi(y)\mid x\in X\land y\in Y} &\text{def.\ of $\Pi$}\\
		&=\set{a+b\mid a\in\Pi(X)\land b\in\Pi(y)}\\
		&=\Pi(X)\dotplus \Pi(Y)&\text{def. of $\dotplus$}
	\end{align*}
\qed
\end{proof}
\fi

\smallskip %
\noindent %
{\bf Context-free Languages.}
A \emph{context-free grammar} $G$ is a tuple $(\mathcal{X},\Sigma,\delta)$
where $\mathcal{X}$ is a finite non-empty set of variables (non-terminal
letters), $\Sigma$ is an alphabet of terminal letters and $\delta \subseteq
\mathcal{X}\times (\Sigma \cup \mathcal{X})^*$ a finite set of productions (the
production $(X,w)$ may also be noted $X\rightarrow w$). Given two strings $u,v
\in (\Sigma \cup \mathcal{X})^*$ we define the relation $u \Rightarrow v$, if
there exists a production $(X, w)\in\delta$ and some words $y,z \in (\Sigma
\cup \mathcal{X})^*$ such that $u=yXz$ and $v=ywz$.
We use $\Rightarrow^*$ for the reflexive transitive closure of $\Rightarrow$.
A word $w\in \Sigma^*$ is recognized by the grammar $G$ from the state $X\in
\mathcal{X}$ if $X\Rightarrow^* w$.  Given $X\in \mathcal{X}$, the language
$L_X(G)$ is given by $\set{w\in\Sigma^*\mid X\Rightarrow^* w}$. A language $L$
is \emph{context-free} (written CFL) if there exists a context-free grammar
$G=(\mathcal{X},\Sigma,\delta)$ and an initial variable $X\in\mathcal{X}$ such
that is $L=L_X(G)$.  A \emph{linear grammar} $G$ is a context-free grammar where
each production is in $\mathcal{X}\times \Sigma^* (\mathcal{X}\cup
\set{\varepsilon}) \Sigma^*$.  A language $L$ is \emph{linear} if $L = L_X(G)$
for some linear grammar $G$ and initial variable $X$ of $G$.  A CFL $L$ is {\em
bounded} if it is a subset of some elementary bounded language.

\noindent{\bf Proof Plan.}
The main result of the paper is the following.

\begin{theorem}\label{theo-main}
For every CFL $L$, there is an effectively
computable CFL $L'$ such that
$(i)$ $L'\subseteq L$,
$(ii)$ $\Pi(L)=\Pi(L')$,
and $(iii)$ $L'$ is bounded.
\end{theorem}%
\noindent
We actually solve the following related problem in our proof.
\begin{problem}
Given a CFL $L$, compute an elementary bounded language $B$
such that $\Pi(L\cap B)=\Pi(L)$.
\label{pb:elembounded}
\end{problem}

If we can compute such a $B$, then we can compute the CFL $L' = B\cap L$
which satisfies conditions $(i)$ to $(iii)$ of the Th.~\ref{theo-main}.
Thus, solving Pb.~\ref{pb:elembounded} proves the theorem constructively.

We solve Pb.~\ref{pb:elembounded} for a language $L$ as follows: 
(1) we find an $L'$
such that $L'\subseteq L$, $\Pi(L') = \Pi(L)$, 
and $L'$ has a ``simple'' structure
(Sect.~\ref{sec-representation}) and (2) then show how to find an elementary bounded
$B$ with $\Pi(L'\cap B) = \Pi(L')$, assuming
this structure (Sect.~\ref{sec-construction}).
Observe that if $L'\subseteq L$ and $\Pi(L)=\Pi(L')$, then
for every elementary bounded $B$, we have $\Pi(L'\cap B) = \Pi(L')$ implies
$\Pi(L\cap B) = \Pi(L)$ as well. 
So the solution $B$ for $L'$ in step (2) is a solution for $L$ as well.
\iflong
Section~\ref{sec-applications} provides applications of the result for program
analysis problems.
\fi
\ifshort
Section~\ref{sec-applications} provides an application of the result for multithreaded program analysis 
and compares it with an existing technique.
\fi

\section{A Parikh-Equivalent Representation}
\label{sec-representation}

Our proof to compute the above $L'$ relies on a fixpoint characterization of CFLs
and their Parikh image. Accordingly, we introduce the necessary mathematical
notions to define and study properties of those fixpoints.

\smallskip %
\noindent %
{\bf Semiring.}
A \emph{semiring} $\ring$ is a tuple $\tuple{S,\comb,\ext,\zero,\one}$, where
$S$ is a set with $\zero,\one \in S$, $\tuple{S,\comb,\zero}$ is a commutative
monoid with neutral element $\zero$, $\tuple{S,\ext,\one}$ is a monoid with
neutral element $\one$, $\zero$ is an annihilator w.r.t. $\ext$, \ie $\zero
\ext a = a \ext \zero = \zero$ for all $a \in S$, and $\ext$ distributes over
$\comb$, \ie $a \ext (b \comb c) = (a \ext b) \comb (a \ext c)$, and $(a \comb
b) \ext c = (a \ext c) \comb (b \ext c)$. We call $\oplus$ the \emph{combine}
operation and $\odot$ the \emph{extend} operation.  The natural order relation
$\sqsubseteq$ on a semiring $\ring$ is defined by $a \sqsubseteq b
\Leftrightarrow \exists d \in S \colon a \comb d = b$. The semiring $\ring$ is
\emph{naturally ordered} if $\sqsubseteq$ is a partial order on $S$.  The
semiring $\mathcal{S}$ is \emph{commutative} if $a\ext b=b\ext a$ for all $a,b
\in S$, \emph{idempotent} if $a\comb a =a$ for all $a\in S$, \emph{complete} if
it is naturally ordered and $\sqsubseteq$ is such that $\omega$-chains
$a_0\sqsubseteq a_1\sqsubseteq \cdots \sqsubseteq a_n \sqsubseteq \cdots$ have
least upper bounds.
Finally, the semiring $\ring$
is \emph{$\omega$-continuous} if it is naturally ordered, complete and for all
sequences $(a_i)_{i\in \nats}$ with $a_i\in S$, $\sup \left\{\bigcomb_{i=0}^n
a_i \mid n\in \nats\right\}=\bigcomb_{i\in \nats} a_i$.  We define two
semirings we shall use subsequently.

\begin{description}
\item[{\it Language Semiring.}]
Let $\lring = \tuple{2^{\Sigma^*},\cup,\cdot,\emptyset,\set{\varepsilon}}$
denote the idempotent $\omega$-continuous semiring of languages.
The natural order on $\lring$ is given by set inclusion (viz. $\subseteq$).
\item[{\it Parikh Semiring.}]
The tuple $\pring = \Bigl\langle 2^{\nats^{p}},\cup,\dotplus,\emptyset,\{\pzero\}\Bigr\rangle$ is
the idempotent $\omega$-continuous commutative semiring of Parikh vectors.
The natural order is again given by $\subseteq$.
\end{description}

\iflong
\smallskip %
\noindent %
{\bf Valuation, partial order, linear form, monomial and polynomial (transformation).}
A \emph{valuation} $\vbf$ is a mapping $\mathcal{X} \rightarrow S$.  We
denote by $\ring^\mathcal{X}$ the set of all valuations and by $\vzero$ the
valuation which maps each variable to $\zero$.\\
The operations $\comb$, $\ext$ are naturally extended to valuations.
The \emph{partial order} $\sqsubseteq$ on
$\ring$ can be lifted to a partial order on valuations, to this end we stack a
point above $\sqsubseteq$ (viz.  $\mathrel{\smash[t]{\stackrel{.}{\sqsubseteq}}}$) to denote the
pointwise inclusion, given by $\vbf\mathrel{\smash[t]{\stackrel{.}{\sqsubseteq}}}\vbf'$ if and
only if $\vbf(X)\sqsubseteq\vbf'(X)$ for every $X\in \mathcal{X}$.  \\ A
\emph{linear form} is a mapping $l\colon \ring^\mathcal{X}\rightarrow S$
satisfying $l(\vbf \comb \vbf') =l(\vbf)\comb l(\vbf')$ for every
$\vbf,\vbf'\in \ring^\mathcal{X}$ and $l(\vzero)=\zero$.  \\ 
A \emph{monomial}
is a mapping $\ring^\mathcal{X} \rightarrow S$ described by a finite expression
$m=a_1 \ext X_1 \ext a_2 \ldots a_k \ext X_k \ext a_{k+1}$ where $k\geq 0$,
$a_1,\ldots,a_{k+1}\in S$ and $X_1,\ldots X_k\in \mathcal{X}$ such that
$m(\vbf)=a_1 \ext \vbf(X_1) \ext a_2 \ldots a_k \ext\vbf(X_k) \ext a_{k+1}$ for 
$\vbf\in \ring^\mathcal{X}$. 
The empty monomial is given by an empty expression
coincides with $\one$. 
\\ 
A \emph{polynomial} is a finite
combination of monomials : $f=m_1 \comb \cdots \comb m_k$ where $k\geq 0$ and
$m_1,\ldots,m_k$ are monomials. The set of
polynomials w.r.t. $\ring$ and $\mathcal{X}$ will be denoted by
$\ring[\mathcal{X}]$.
The empty polynomial is given by an empty combination of monomials and
coincides with $\zero$.
\\
Finally, a \emph{polynomial transformation} $\funbf$ is a mapping
$\ring^\mathcal{X} \rightarrow \ring^\mathcal{X}$ described by the set
$\set{\funbf_X\in\ring[\mathcal{X}] \mid X\in \mathcal{X}}$ of polynomials:
hence, for every valuation $\vbf\in \ring^\mathcal{X}$, $\funbf(\vbf)$ is a valuation that assigns each variable $X\in\mathcal{X}$ to $\funbf_X(\vbf)$.

\smallskip %
\noindent %
{\bf Differential.}
For every $X\in \mathcal{X}$, let $\mathrm{d}X$ denote the linear form defined
by $\mathrm{d}X(\vbf)=\vbf(X)$ for every $\vbf\in \ring^{\mathcal{X}}$:
$\mathrm{d}X$ is the \emph{dual variable} associated with the variable $X$.
Let $\mathrm{d}\mathcal{X}$ denote the set $\set{\mathrm{d}X\mid X\in
\mathcal{X}}$ of dual variables.

Let $f\in \ring[\mathcal{X}]$ be a polynomial and let $X\in \mathcal{X}$ be a
variable.  The \emph{differential w.r.t. $X$} of $f$ is the mapping $D_X
f\colon \ring^\mathcal{X} \rightarrow \ring^\mathcal{X} \rightarrow S$ that
assigns to every valuation $\vbf$ the linear form $D_X \deriv{f}_{\vbf}$ defined
by induction as follows:
\[D_X \deriv{f}_{\vbf}=
	\begin{cases}
	\zero & \text{ if $f\in S$  or $f\in \mathcal{X}\setminus \set{X}$}\\
	\mathrm{d}X & \text{ if  $f=X$}\\
	D_X \deriv{g}_{\vbf} \ext h(\vbf) \comb g(\vbf)\ext D_X \deriv{h}_{\vbf}
	& \text{ if  $f=g\ext h$} \\
	D_X \deriv{g}_{\vbf} \comb D_X \deriv{h}_{\vbf} & \text{ if $f=g\comb h$ \enspace .}
\end{cases} \]
Then, the \emph{differential} of $f$ is defined by \[Df = \bigcomb_{X\in
\mathcal{X}} D_X f \enspace .\] Consequently, the linear form $D
\deriv{f}_{\vbf}$ is a polynomial of the following form: \[ (a_1 \ext
\mathrm{d}X_1 \ext a'_1) \comb \dots \comb (a_k \ext \mathrm{d}X_k
\ext a'_k) \] where each $a_i,a'_i\in S$ and $X_i\in \mathcal{X}$.  We
extend the definition of differential on polynomial transformation. Hence,
$D\funbf:\ring^\mathcal{X} \rightarrow \ring^\mathcal{X} \rightarrow
\ring^\mathcal{X}$ is defined for every $\vbf,\wbf \in \ring^\mathcal{X}$ and
every variable $X$ as follows: \[(D \deriv{\funbf}_{\vbf} (\wbf))(X)=D
\deriv{\funbf_X}_{\vbf} (\wbf) \enspace .\]

\noindent %
{\bf Least Fixpoint.}
Recall that a mapping $f \colon \ring \rightarrow \ring$ is monotone if $a\sqsubseteq b$
implies $f(a)\sqsubseteq f(b)$, and continuous if for any infinite
chain $a_0, a_1, a_2,\dots$ we have $\mathrm{sup}\set{f(a_i)} =
f(\mathrm{sup}\set{a_i})$. The definition can be extended to mappings $\funbf
\colon \ring^\mathcal{X} \rightarrow \ring^\mathcal{X}$ from valuations to valuations in the obvious way (componentwise).
Then we may formulate the following proposition (cf. \cite{kuich}).

\begin{proposition}
Let $\funbf$ be a polynomial transformation. The mapping induced by $\funbf$ is
monotone and continuous. Hence, by Kleene's theorem, $\funbf$ has a unique
least fixpoint $\mu\funbf$. Further, $\mu\funbf$ is the supremum (w.r.t.
$\stackrel{\cdot}{\sqsubseteq}$) of the Kleene's iteration sequence given by $\boldsymbol{\eta}_0 = \funbf(\vzero)$, and
$\boldsymbol{\eta}_{i+1}= \funbf(\boldsymbol{\eta}_i)$.
\end{proposition}
\fi

\ifshort 
\smallskip %
\noindent %
{\bf Valuation, polynomial.}
In what follows, let $\mathcal{X}$ be a finite set of variables and $\ring=\tuple{S,\comb,\ext,\zero,\one}$ be
an $\omega$-continuous semiring.

A \emph{valuation} $\vbf$ is a mapping $\mathcal{X} \rightarrow S$.  We denote
by $\ring^\mathcal{X}$ the set of all valuations and by $\vzero$ the valuation
which maps each variable to $\zero$.  We define
$\mathrel{\smash[t]{\stackrel{.}{\sqsubseteq}}}\subseteq \ring^\mathcal{X}\times
\ring^\mathcal{X}$ as the order given by
$\vbf\mathrel{\smash[t]{\stackrel{.}{\sqsubseteq}}}\vbf'$ if and only if
$\vbf(X)\sqsubseteq\vbf'(X)$ for every $X\in \mathcal{X}$.
A \emph{monomial} is a mapping $\ring^\mathcal{X} \rightarrow S$ given by a
finite expression $m=a_1 \ext X_1 \ext a_2 \cdots a_{k} \ext X_{k} \ext a_{k+1}$
where $k\geq 0$, $a_1,\ldots,a_{k+1}\in S$ and $X_1,\ldots, X_k\in \mathcal{X}$
such that $m(\vbf)=a_1 \ext \vbf(X_1) \ext a_2 \cdots a_k \ext\vbf(X_k) \ext
a_{k+1}$ for $\vbf\in \ring^\mathcal{X}$.  \\ A \emph{polynomial} is a finite
combination of monomials : $f=m_1 \comb \cdots \comb m_k$ where $k\geq 0$ and
$m_1,\ldots,m_k$ are monomials. The set of polynomials w.r.t. $\ring$ and
$\mathcal{X}$ will be denoted by $\ring[\mathcal{X}]$.  Finally, a
\emph{polynomial transformation} $\funbf$ is a mapping $\ring^\mathcal{X}
\rightarrow \ring^\mathcal{X}$ described by the set
$\set{\funbf_X\in\ring[\mathcal{X}] \mid X\in \mathcal{X}}$ of polynomials :
hence, for every vector $\vbf\in \ring^\mathcal{X}$, $\funbf(\vbf)$ is a 
valuation of each variable $X\in\mathcal{X}$ to $\funbf_X(\vbf)$.

\noindent %
{\bf Least Fixpoint.}
Recall that a mapping $f \colon S \rightarrow S$ is monotone if $a\sqsubseteq
b$ implies $f(a)\sqsubseteq f(b)$, and continuous if for any infinite chain
$a_0, a_1, a_2,\dots$ we have $\mathrm{sup}\set{f(a_i)} =
f(\mathrm{sup}\set{a_i})$. The definition can be extended to mappings $\funbf
\colon \ring^\mathcal{X} \rightarrow \ring^\mathcal{X}$ in the obvious way
(using $\mathrel{\smash[t]{\stackrel{.}{\sqsubseteq}}}$).  Then we may formulate the
following proposition (cf. \cite{kuich}).

\begin{proposition}
Let $\funbf$ be a polynomial transformation. The mapping induced by $\funbf$ is
monotone and continuous and $\funbf$ has a unique least fixpoint $\mu\funbf$.
\end{proposition}
\fi

Fixpoints of polynomial transformations relates to CFLs as follows.
Given a grammar $G=(\mathcal{X},\Sigma,\delta)$, let $L(G)$ be the valuation
which maps each variable $X\in\mathcal{X}$ to the language $L_X(G)$. We first
characterize the valuation $L(G)$ as the least fixpoint of a polynomial
transformation $\funbf$ defined as follows:
each $\funbf_{X}$ of $\funbf$ is
given by the combination of $\alpha$'s for $(X,\alpha)\in\delta$ where $\alpha$
is interpreted as a monomial on the semiring $\lring$.
From \cite{ChomskySchutzenberger} we know that $L(G)=\mu\funbf$.

\begin{example}
Let $G=(\set{X_0,X_1},\set{a,b},\delta)$
where $\delta=\{(X_0 \rightarrow a X_1 \vert a),\, (X_1 \rightarrow X_0 b \vert a X_1 b X_0)\}$.
It defines the polynomial transformation $\funbf$ on $\lring^{\mathcal{X}}$ such
that $\funbf_{X_0}=a\cdot X_1\cup a$ and $\funbf_{X_1}=X_0\cdot b \cup a\cdot X_1\cdot b \cdot X_0$, and $L(G)$ is the
least fixpoint of $\funbf$ in the language semiring.
\qed
\end{example}

\iflong
We now recall the iteration sequence of \cite{EKL07:dlt,EKL07} whose limit is
the least fixpoint of $\funbf$. In some cases, the iteration sequence
converges after a finite number of iterates while the Kleene iteration sequence
does not.

\smallskip %
\noindent %
{\bf Newton's Iteration Sequence.}
Given a polynomial transformation $\funbf$ on a $\omega$-continuous semiring
$\ring$, \emph{Newton's iteration sequence} is given by the following
sequence: \[\bmu_{0}=\funbf(\vzero)\qquad \text{and}\qquad
\bmu_{i+1}=D\deriv{\funbf}_{\bmu_{i}}^*(\funbf(\bmu_{i}))\] the limit of which
coincides with $\mu\funbf$ (see \cite{EKL07,EKL07:dlt} for further details).
\fi

\subsection{Relating the Semirings}

\iflong
We naturally extend the definition of the Parikh image to a valuation $\vbf\in \lring^\mathcal{X}$ as the valuation of $\pring^\mathcal{X}$ defined for each variable $X$ by: $\Pi(\vbf)(X)=\Pi(\vbf(X))$. The following lemma relates polynomial transformations on $\lring$ and $\pring$.
\begin{lemma}
	Let $f_{\lring}\in\lring[\mathcal{X}]$, that is a polynomial over the semiring $\lring$ and variables $\mathcal{X}$. Define
	$f_{\pring}=\Pi\comp f_{\lring}\comp \Pi^{-1}$, we have
	$f_{\pring}\in\pring[\mathcal{X}]$.
	\label{lem:fppoly}
\end{lemma}
\begin{proof}
	By induction on the structure of $f_{\lring}$. The polynomial $f_{\lring}$ is given by $m_1 \cup \dots \cup m_{\ell}$. Hence,
\begin{align*}
	\Pi\comp f_{\lring}\comp \Pi^{-1}&=\Pi \comp (m_1 \cup \dots \cup m_{\ell})\comp\Pi^{-1}\\
	&=\Pi\comp m_1\comp\Pi^{-1}\cup\dots\cup\Pi\comp m_{\ell}\comp\Pi^{-1}
\end{align*}
where each $m_i$ is of the form $a_1\cdot X_1\cdot a_2\ldots a_k\cdot X_k\cdot a_{k+1}$ with
$a_1,\dots,a_{k+1}\subseteq\Sigma^*$, $X_1,\dots,X_k\in\mathcal{X}$.
Let $m$ be a monomial, we have:
\begin{align*}
	\Pi\comp m\comp\Pi^{-1}&=\Pi\comp a_1\cdot X_1\cdot a_2\ldots a_k\cdot X_k\cdot a_{k+1}\comp\Pi^{-1}\\
	&=\Pi(a_1)\dotplus X_1\dotplus \Pi(a_2)\ldots \Pi(a_k)\dotplus X_k\dotplus \Pi(a_{k+1}) &\text{id.\ of $\Pi\comp\Pi^{-1}$, preser.\ of $\Pi$}
\end{align*}
\qed
\end{proof}

We now prove a commutativity results on polynomials and the Parikh mapping.

\begin{lemma}
	Let $f_{\lring}\in\lring[\mathcal{X}]$,
	for every valuation $\vbf\in \lring^{\mathcal{X}}$, we have:
	\[\Pi(f_{\lring}(\vbf))=f_{\pring}(\Pi(\vbf))\enspace .\]
	\label{lem:piffip}
\end{lemma}
\begin{proof}
First, the definition of $f_{\pring}$ shows that for every $\vbf\in \lring^{\mathcal{X}}$: 
\begin{align*}
&\Pi\comp f_{\lring}(\vbf)=f_{\pring}\comp \Pi(\vbf)\\
&\text{if{}f }\Pi\comp f_{\lring}(\vbf)=\Pi\comp f_{\lring} \comp \Pi^{-1}\comp \Pi(\vbf)\\
&\text{only if }\Pi^{-1}\comp\Pi\comp f_{\lring}(\vbf)=\Pi^{-1}\comp\Pi\comp f_{\lring} \comp \Pi^{-1}\comp \Pi(\vbf)&\text{appl.\ of $\Pi^{-1}$}
\intertext{Moreover, }
&\Pi^{-1}\comp\Pi\comp f_{\lring}(\vbf)=\Pi^{-1}\comp\Pi\comp f_{\lring} \comp \Pi^{-1}\comp \Pi(\vbf)\\
&\text{only if }\Pi\comp\Pi^{-1}\comp\Pi\comp f_{\lring}(\vbf)=\Pi\comp\Pi^{-1}\comp\Pi\comp f_{\lring} \comp \Pi^{-1}\comp \Pi(\vbf)&\text{appl.\ of $\Pi$}\\
&\text{only if } \Pi\comp f_{\lring}(\vbf)=\Pi\comp f_{\lring} \comp \Pi^{-1}\comp \Pi(\vbf) &\text{identity of $\Pi\comp\Pi^{-1}$}\\
&\text{if{}f } \Pi\comp f_{\lring}(\vbf)=f_{\pring}\comp \Pi(\vbf)&\text{def.\ of $f_{\pring}$}
\intertext{Hence, }
&\Pi\comp f_{\lring}(\vbf)=f_{\pring}\comp\Pi(\vbf)\text{ if{}f }\Pi^{-1}\comp\Pi\comp f_{\lring}(\vbf)=\Pi^{-1}\comp\Pi\comp f_{\lring} \comp \Pi^{-1}\comp \Pi(\vbf)
\intertext{Let $\phi=\Pi^{-1}\comp\Pi$, we will thus show that for every $\vbf\in\lring^{\mathcal{X}}$}
&\phi\comp f_{\lring}(\vbf)=\phi\comp f_{\lring} \comp \phi(\vbf)
\end{align*}
The inclusion $\phi\comp f_{\lring}(\vbf)\subseteq\phi\comp f_{\lring} \comp
\phi(\vbf)$ is clear since $\vbf\mathrel{\smash[t]{\stackrel{.}{\sqsubseteq}}}\phi(\vbf)$, every function occuring in the
above expression is monotone and the functional composition preserves
monotonicity.  For the reverse inclusion, we first show that for every
$\wbf\mathrel{\smash[t]{\stackrel{.}{\sqsubseteq}}}\phi(\vbf)$ we have $f_{\lring}(\wbf)\subseteq \phi\comp
f_{\lring}(\vbf)$.  That is $\forall x\in f_{\lring}(\wbf)\colon x\in\phi\comp
f_{\lring}(\vbf)$.  $f_{\lring}\in\lring[\mathcal{X}]$ shows that $x\in m(\wbf)$ for some monomial $m=a_1\cdot X_1\cdot a_2\ldots a_k\cdot X_k\cdot
a_{k+1}$, that is $x\in a_1\cdot \wbf(X_1)\cdot a_2\ldots a_k\cdot
\wbf(X_k)\cdot a_{k+1}$.  We have,
\begin{align*}
	\phi\comp f_{\lring}(\vbf)&\supseteq \phi(a_1\cdot \vbf(X_1)\cdot a_2\ldots a_k\cdot \vbf(X_k)\cdot a_{k+1})\\
	&\supseteq \phi(a_1)\cdot \phi(\vbf(X_1))\cdot \phi(a_2)\ldots \phi(a_k)\cdot \phi(\vbf(X_k))\cdot \phi(a_{k+1})&\text{struct.\ semipreserv.}\\
	&\supseteq a_1\cdot \phi(\vbf(X_1))\cdot a_2\ldots a_k\cdot \phi(\vbf(X_k))\cdot a_{k+1}&\text{extensivity of $\phi$}\\
	&\supseteq a_1\cdot \wbf(X_1)\cdot a_2\ldots a_k\cdot \wbf(X_k)\cdot a_{k+1} &\text{$\wbf\mathrel{\smash[t]{\stackrel{.}{\sqsubseteq}}}\phi(\vbf)$}\\
	&\owns x &\text{def.\ of $x$}
\end{align*}
The following reasoning concludes the proof:
\begin{align*}
&f_{\lring}\comp\phi(\vbf)\subseteq \phi\comp f_{\lring}(\vbf)&\text{from
above with } \wbf=\phi(\vbf)\\
&\text{only if }\phi\comp f_{\lring}\comp\phi(\vbf)\subseteq \phi\comp\phi\comp f_{\lring}(\vbf)&\text{monotonicity of $\phi$}\\
&\text{if{}f }\phi\comp f_{\lring}\comp\phi(\vbf)\subseteq \phi\comp f_{\lring}(\vbf)&\text{idempotency of $\phi$}
\end{align*}
\qed
\end{proof}

Here follows a commutativity result between the differential and the Parikh image.

\begin{lemma}
	For every $f_{\lring}\in\lring[\mathcal{X}]$, every valuation $\vbf,\wbf\in \lring^{\mathcal{X}}$, every $X\in \mathcal{X}$ we have:
	\[\Pi(D_X\deriv{f_{\lring}}_{\vbf}(\wbf))=D_X\deriv{f_{\pring}}_{\Pi(\vbf)}(\Pi(\wbf))\enspace .\]
	\label{lem:deriv}
\end{lemma}
\begin{proof}
	First it is important to note that Lemma~\ref{lem:fppoly} shows that
	$f_{\pring}$ and $f_{\lring}$ are of the same form. Then the proof
	falls into four parts according to the definition of the differential w.r.t. $X$.
	\medskip

	\noindent{$f_{\lring}\in2^{\Sigma^*}$ or $f_{\lring}\in \mathcal{X}\setminus\set{X}$.} In
	this case, we find that $D_X\deriv{f_{\lring}}_{\vbf}(\wbf)=\emptyset$,
	hence that $\Pi(D_X\deriv{f_{\lring}}_{\vbf}(\wbf))=\emptyset$. 
	Since $f_{\pring}$ is of the above form, we find that
	$D_X\deriv{f_{\pring}}_{\Pi(\vbf)}(\Pi(\wbf))=\emptyset$.\medskip

	\noindent{$f_{\lring}=X$.} So $f_{\pring}=X$.
	\begin{align*}
		\Pi(D_X\deriv{X}_{\vbf}(\wbf))&=\Pi(\mathrm{d}X(\wbf))&\text{def.\ of diff}\\
		&=\Pi(\wbf(X))&\text{def.\ of $\mathrm{d}X$}\\
		&=\Pi(\wbf)(X)&\text{def.\ of $\Pi$}\\
		&=\mathrm{d}X(\Pi(\wbf))&\text{def.\ of $\mathrm{d}X$}\\
		&=D_X\deriv{X}_{\Pi(\vbf)}(\Pi(\wbf))&\text{def.\ of diff}
	\end{align*}

	\noindent{$f_{\lring}=g_{\lring}\cdot h_{\lring}$} So $f_{\pring}$ is
	of the form $g_{\pring}\dotplus h_{\pring}$. The induction hypothesis
	shows the rest.

	\noindent{$f_{\lring}=\bigcup_{i\in I}f_i$} this case is treated similarly.
	\qed
\end{proof}

This result generalizes to the complete differential :
\[\Pi(D\deriv{f_{\lring}}_{\vbf}(\wbf))=D\deriv{f_{\pring}}_{\Pi(\vbf)}(\Pi(\wbf))\enspace .\]

We note that the previous results also generalizes to polynomial transformation
in a natural way. In the next subsection, thanks to the previous results, we
show that Newton's iteration sequence on the language semiring reaches a
stable Parikh image after a finite number of steps. This result is crucial in
order to achieve the goal of this section: compute a sublanguage $L'$ of $L$
such that $\Pi(L)=\Pi(L')$.

\subsection{Convergence of Newton's Iteration}
\fi

Given a polynomial transformation $\funbf$, we now characterize the
relationship between the least fixpoints $\mu \funbf$ taken over the language
and the Parikh semiring, respectively. Either fixpoint is given by the limit of
a sequence of \emph{iterates} which is defined by Newton's iteration scheme
\cite{EKL07:dlt,EKL07}.  Our characterization operates at the level of those
iterates: we inductively relate the iterates of each iteration sequence (over
the Parikh and language semirings).  We use Newton's iteration instead of the
usual Kleene's iteration sequence because Newton's iteration is guaranteed to
converge on the Parikh semiring in a finite number of steps, a property that we
shall exploit.  Kleene's iteration sequence, on the other hand, may not
converge. 
\ifshort
Due to lack of space, we refer the reader to \cite{EKL07:dlt,EKL07} for Newton's iteration scheme definition.

We first extend the definition of the Parikh image to a valuation $\vbf\in
\lring^\mathcal{X}$ as the valuation of $\pring^\mathcal{X}$ defined for each
variable $X$ by: $\Pi(\vbf)(X)=\Pi(\vbf(X))$.  
Then, given
$\funbf_{\lring}\colon \lring^{\mathcal{X}}\rightarrow \lring^{\mathcal{X}}$, a
polynomial transformation, we define a polynomial transformation
$\funbf_{\pring}\colon\pring^{\mathcal{X}}\rightarrow\pring^{\mathcal{X}}$ such
that:  for every $X\in\mathcal{X}$ we have ${\funbf_{\pring}}_X=
\Pi\comp{\funbf_{\lring}}_X\comp\Pi^{-1}$.  
\fi
Lemma.~\ref{lem:parikhcoincidence} relates the iterates
for $\mu\funbf_{\lring}$ and $\mu\funbf_{\pring}$ using the Parikh image
mapping.

\begin{lemma}
	Let $(\bnu_{i})_{i\in\nats}$ and $(\bkappa_{i})_{i\in\nats}$
	be Newton's iteration sequences associated with $\funbf_{\lring}$ and $\funbf_{\pring}$, respectively.
	For every $i\in\nats$, we have $\Pi(\bnu_{i})=\bkappa_{i}$.
	\label{lem:parikhcoincidence}
\end{lemma}
\iflong
\begin{proof}
	\noindent {\bf base case. $(i=0)$} This case is trivially solved
	using part (2) of Lem.~\ref{lem:fppoly}.

  \smallskip
	\noindent {\bf inductive case. $(i+1)$}
	\begin{align*}
		\Pi(\bnu_{i+1})&=\Pi(D\deriv{\funbf_{\lring}}_{\bnu_i}^{*}(\funbf_{\lring}(\bnu_i)))\\
		&=\Pi(\bigcup_{j\in\nats} D\deriv{\funbf_{\lring}}_{\bnu_i}^{j}(\funbf_{\lring}(\bnu_i)))&\text{def.\ of ${}^*$}\\
		&=\bigcup_{j\in\nats} \Pi(D\deriv{\funbf_{\lring}}_{\bnu_i}^{j}(\funbf_{\lring}(\bnu_i)))&\text{additivity of $\Pi$}\\
		&=\bigcup_{j\in\nats} \Pi\bigl(D\deriv{\funbf_{\lring}}_{\bnu_i}(D\deriv{\funbf_{\lring}}_{\bnu_i}^{j-1}(\funbf_{\lring}(\bnu_i)))\bigr)&\text{funct.\ comp.}\\
		&=\bigcup_{j\in\nats} D\deriv{\funbf_{\pring}}_{\Pi(\bnu_i)}\bigl(\Pi(D\deriv{\funbf_{\lring}}
		_{\bnu_i}^{j-1}(\funbf_{\lring}(\bnu_i)))\bigr) & \text{Lem.~\ref{lem:deriv}}\\
		&=\bigcup_{j\in\nats} D\deriv{\funbf_{\pring}}_{\Pi(\bnu_i)}^{j}(\Pi(\funbf_{\lring}(\bnu_i))) & \text{$j-1\times$ Lem.~\ref{lem:deriv}}\\
		&=\bigcup_{j\in\nats} D\deriv{\funbf_{\pring}}_{\Pi(\bnu_i)}^{j}(\funbf_{\pring}(\Pi(\bnu_i))) & \text{Lem.~\ref{lem:fppoly}}\\
		&=\bigcup_{j\in\nats} D\deriv{\funbf_{\pring}}_{\bkappa_i}^{j}(\funbf_{\pring}(\bkappa_i)) & \text{ind.\ hyp.}\\
		&= D\deriv{\funbf_{\pring}}_{\bkappa_i}^{*}(\funbf_{\pring}(\bkappa_i))\\
		&=\bkappa_{i+1}
	\end{align*}
\qed
\end{proof}
\fi

In \cite{EKL07}, the authors show that Newton's iterates converges after a
finite number of steps when defined over a commutative $\omega$-continuous
semiring. This shows, in our setting, that $(\bkappa_{i})_{i\in\nats}$ stabilizes after a finite number of steps.
\begin{lemma}
	Let $(\bkappa_{i})_{i\in\nats}$ be Newton's iteration sequence
	associated to $\funbf_{\pring}$ and let $n$ be the number of variables in
	$\mathcal{X}$. For every $k\geq n$, we have
	$\bkappa_k=\Pi(\mu \funbf_{\lring})$.
	Hence, for every $k\geq n$, $\Pi(\bnu_k)=\Pi(\mu \funbf_{\lring})$.
\label{lem:newtonparikh}
\end{lemma}
\iflong
\begin{proof}
	\begin{align*}
		&\bkappa_i=\Pi(\bnu_i) &\text{for each $i\in\nats$ by Lem.~\ref{lem:parikhcoincidence}}\\
		&\Rightarrow\bigcup_{i\in\nats}\bkappa_i=\bigcup_{i\in\nats}\Pi(\bnu_i)\\
		&\Leftrightarrow \mu \funbf_{\pring}=\bigcup_{i\in\nats} \Pi(\bnu_i)&
		\text{$\omega$-continuity of $\pring$}\\
		&\Leftrightarrow \mu \funbf_{\pring}=\Pi(\bigcup_{i\in\nats} \bnu_i)&
		\text{additivity of $\Pi$}\\
		&\Leftrightarrow \mu \funbf_{\pring}=\Pi(\mu \funbf_{\lring})&
		\text{$\omega$-continuity of $\lring$}\\
		&\Rightarrow \bkappa_k=\Pi(\mu \funbf_{\lring})&
		\text{for every $k\geq n$ by Th.~6 of \cite{EKL07}}
	\end{align*}
	Transitivity of the equality shows the remaining result.
\qed
\end{proof}
\fi

We know Newton's iteration sequence $(\bnu_i)_{i\in\mathbb{N}}$, whose limit is
$\mu\funbf_{\lring}$, may not converge after a finite number of iterations.
However, using Lem.~\ref{lem:newtonparikh}, we know that the Parikh image of
the iterates stabilizes after a finite number of steps.
Precisely, 
if $n$ is the number of variables in $\mathcal{X}$,
then the language given by $\bnu_n$ is such that $\Pi(\bnu_n)=\Pi(L(G))$.
Moreover because $(\bnu_i)_{i\in\mathbb{N}}$ is an ascending chain, for each
variable $X\in\mathcal{X}$, we have that $\bnu_n(X)$ is a sublanguage of
$L_{X}(G)$ such that $\Pi(\bnu_n(X))=\Pi(L_{X}(G))$.

\subsection{Representation of Iterates}
\label{subsec:representation}

We now show that Newton's iterates can be effectively represented as a
combination of linear grammars and homomorphisms.

A \emph{substitution} $\sigma$ from alphabet $\Sigma_1$ to alphabet $\Sigma_2$
is a function which maps every word over $\Sigma_1$ to a set of words of
$\Sigma_2^*$ such that $\sigma(\varepsilon)=\set{\varepsilon}$ and $\sigma(u
\cdot v)=\sigma(u)\cdot \sigma(v)$.  A \emph{homomorphism} $h$ is a
substitution such that for each word $u$, $h(u)$ is a singleton.  We define the
substitution $\sigma_{[a/b]}\colon \Sigma_1\cup\set{a}\rightarrow
\Sigma_1\cup\set{b}$ which maps $a$ to $b$ and leaves all other symbols
unchanged.\\
We show below that the iterates $(\bnu_k)_{k\leq n}$ have a ``nice'' representation.

\iflong
Let us leave for a moment Newton's iteration sequence and
turn to our initial problem as stated in Pb.~\ref{pb:elembounded}.  Let
$L$ be a context-free language, our goal is to compute a
sublanguage $L'$ such that $\Pi(L)=\Pi(L')$ (then we solve Pb.~\ref{pb:elembounded} on instance
$L'$ instead of $L$ because it is equivalent). Below we give an effective procedure to compute such a $L'$ based on the
previously defined iteration sequences and the convergence results.


Given a grammar $G=(\mathcal{X},\Sigma,\delta)$, let $L(G)$ be the valuation
which maps each variable $X\in\mathcal{X}$ to the language $L_X(G)$. We first
characterize the valuation $L(G)$ as the least fixpoint of a polynomial
transformation $\funbf$ which is defined using $G$ as follows:
each $\funbf_{X}$ of $\funbf$ is
given by the combination of $\alpha$'s for $(X,\alpha)\in\delta$ where $\alpha$
is now interpreted as a monomial on the semiring $\lring$.

\begin{example}
Let $G=(\set{X_0,X_1},\set{a,b},\delta)$ be the context-free grammar with the production:
\begin{align*}
X_0 &\rightarrow a X_1 \mid a \\
X_1 &\rightarrow X_0 b \mid a X_1 b X_0
\end{align*}
It defines the following polynomial transformation on $\lring^{\mathcal{X}}$
where $\funbf_{X_0}=a X_1 \cup a$ and $\funbf_{X_1}=X_0 b \cup a X_1 b X_0$.
\end{example}

It is well known that $L(G)=\mu\funbf$ (see for instance
\cite{EKL07:dlt}). To evaluate $\mu\funbf$ one can evaluate 
Newton's iteration sequence $(\bnu_i)_{i\in\nats}$ for $\funbf$.
However, a transfinite number of iterates may be needed before reaching
$\mu\funbf$.  We now observe that, by the result of
Lem.~\ref{lem:newtonparikh}, if we consider the iteration sequence
$(\bnu_k)_{k\leq n}$ up to iterate $n$ where $n$ equals to the number of
variables in $\mathcal{X}$ then the language given by $\bnu_n$ is such that
$\Pi(\bnu_n)=\Pi(L(G))$.  Moreover because $(\bnu_i)_{i\in\nats}$ is an
ascending chain we find that: for each variable $X_0\in\mathcal{X}$,
$\bnu_n(X_0)$ is a sublanguage of $L_{X_0}(G)$ such that
$\Pi(\bnu_n(X_0))=\Pi(L_{X_0}(G))$. 

We now explain how to turn this theoretical result into an effective procedure.
Our first step is to define an effective representation for the iterates
$(\bnu_k)_{k\leq n}$.  Our definition is based on the one that was
informally introduced in Example 3.1, part (2) of \cite{EKL07:dlt}. To this
end, we start by defining how to represent the differential
$D\deriv{\funbf}_{\vbf}^*(\funbf(\vbf))$ used in the definition of Newton's
iteration sequence as the language generated by a linear grammar. \medskip

We define $\vbf$ to be the valuation which maps each variable $X\in\mathcal{X}$
to $v_X$ where $v_X$ is a new symbol w.r.t. $\Sigma$. We first observe that $D\deriv{\funbf}_{\vbf}$ is a polynomial
transformation on the set of dual variables $\mathrm{d}\mathcal{X}$
such that the linear form associated to $X$ is a polynomial of the form:
\[ (a_1 \cdot \mathrm{d}X_1
\cdot a'_1) \cup \dots \cup (a_k \cdot \mathrm{d}X_k \cdot a'_k) \] where each
$a_i,a'_i\in(\Sigma\cup\set{v_Y \mid Y\in \mathcal{X}})^*$ and $X_i\in
\mathcal{X}$.  Moreover, $\funbf_{X}$ is a sum of monomials
$m_1,\ldots,m_\ell$.  Hence, we define the linear grammar
$\tilde{G}=(\mathcal{X},\Sigma\cup\set{v_X \mid X\in
\mathcal{X}},\tilde{\delta})$.  For the variable $X$, the set of productions
$\tilde{\delta}$ is: 
\begin{align*}
	X&\rightarrow a_1 X_1 a'_1 \mid \ldots \mid a_k X_k a'_k \\ 
	X&\rightarrow m_1(\vbf) \mid \ldots \mid m_\ell(\vbf) 
\end{align*}
We are able to prove that:
\begin{lemma}
Let $\vbf$ be the valuation which maps each variable $X\in \mathcal{X}$ to $v_X$:
	\[L(\tilde{G})=D\deriv{\funbf}_{\vbf}^*(\funbf(\vbf))\enspace .\]
	\label{lem:lingcoincideiterates}
\end{lemma}
\begin{proof}
	We show by induction the following equivalence. Let $X\in\mathcal{X}$, $w\in\Sigma\cup\set{v_Y\mid Y\in\mathcal{X}}^*$:
	\[ X\Rightarrow^{k+1} w \text{ if{}f } w\in D\deriv{\funbf}_{\vbf}^k(\funbf(\vbf))(X)\enspace .\]

	{\bf Base case. $(k=0)$} In this case, the following equivalence has to be established:	
	\begin{align*}
	&X\Rightarrow w\\
	&\text{if{}f }w\in L_X(X\rightarrow m_1(\vbf) \mid \dots \mid m_\ell(\vbf))\\
	&\text{if{}f }w\in m_1(\vbf) \cup \dots \cup m_\ell(\vbf) &\text{the monomials for $\funbf_X(\vbf)$}\\
	&\text{if{}f }w\in\funbf_X(\vbf)\\
	&\text{if{}f } w\in \funbf(\vbf)(X)
	\end{align*}

	{\bf Inductive case. $(k+1)$}
	\begin{align*}
		&w\in D\deriv{\funbf}_{\vbf}^{k+1}(\funbf(\vbf))(X)\\
		&\text{if{}f }w\in D\deriv{\funbf}_{\vbf}(D\deriv{\funbf}_{\vbf}^{k}(\funbf(\vbf)))(X)&\text{funct.\ comp.}\\
		&\text{if{}f }w\in D\deriv{\funbf_X}_{\vbf}(D\deriv{\funbf}_{\vbf}^{k}(\funbf(\vbf)))\\
		&\text{if{}f }w\in (a_1\cdot dX_1\cdot a'_1) \cup \ldots \cup (a_k \cdot dX_k\cdot a'_k)(D\deriv{\funbf}_{\vbf}^{k}(\funbf(\vbf)))&\text{def.\ of diff.}\\
		&\text{if{}f }\exists i\colon w\in (a_i\cdot dX_i\cdot a'_i)(D\deriv{\funbf}_{\vbf}^{k}(\funbf(\vbf)))\\
		&\text{if{}f }\exists i\,\exists w'\in D\deriv{\funbf}_{\vbf}^{k}(\funbf(\vbf))(X_i) \colon w=a_i\cdot w'\cdot a'_i\\
		&\text{if{}f }\exists i\,\exists w'\colon X\rightarrow a_i X_i a'_i \in \tilde{\delta} \land X_i\Rightarrow^{k+1} w'\land a_i w' a'_i=w\\
		&\text{if{}f } X\Rightarrow^{k+2}w
	\end{align*}
	\qed
\end{proof}

\begin{example}{(cont'd from the previous example)}
The differential of $\funbf$ is given by:
\[D \deriv{\funbf}_{\vbf}=\left(
\begin{array}{c}
a \, \mathrm{d}X_1 \\
\mathrm{d}X_0 \,  b \cup a \, \mathrm{d}X_1 \,  b \, \vbf(X_0) \cup a \,  \vbf(X_1) \,  b \, \mathrm{d}X_0
\end{array}  \right)\]

The grammar $\tilde{G}$ is given by 
$(\set{X_0,X_1},\set{a,b,v_{X_0},v_{X_1}},\tilde{\delta})$
where $\tilde{\delta}$ is such that:
\begin{align*}
	&X_0 \rightarrow a X_1 \mid a v_{X_1} \mid a \\
	&X_1 \rightarrow X_0 b \mid a X_1 b v_{X_0} \mid a v_{X_1} b X_0 \mid v_{X_0} b \mid a v_{X_1} b v_{X_0}\enspace .
\end{align*}
\end{example}
\fi

\noindent %
{\bf $k$-fold composition.}
We effectively compute and represent each iterate as the
valuation which maps each variable $X$ to the language generated by a
$k$-fold composition of a substitution. Since the
substitution maps each symbol onto a language which is linear, it is
effectively represented and manipulated as a linear grammar.
To formally define the representation we need to introduce the following definitions.

Let $\tilde{G}=(\mathcal{X},\Sigma\cup\set{v_X\mid
X\in\mathcal{X}},\tilde{\delta})$ be a linear grammar and let $k\in\nats$,
define $v^k_{\mathcal{X}}$ to be the set of symbols $\set{v^k_X\mid
X\in\mathcal{X}}$.  Given a language $L$ on alphabet $\Sigma\cup\set{v_X\mid
X\in\mathcal{X}}$, we define $L[v_\mathcal{X}^k]$ to be
$\sigma_{[v_X/v_X^k]_{X\in \mathcal{X}}}(L)$. 

For $k\in\set{1,\ldots,n}$, we define $\sigma_k\colon
\Sigma\cup v^{k}_\mathcal{X}\rightarrow
\Sigma\cup v^{k-1}_\mathcal{X}$ as the substitution which maps
each $v_X^k$ onto $L_X(\tilde{G})[v_\mathcal{X}^{k-1}]$ and leaves
$\Sigma$ unchanged. For $k=0$ the substitution $\sigma_0$ maps
each $v_X^0$ on $\funbf(\vzero)(X)$ and leaves $\Sigma$ unchanged.
Let $k,\ell$ be such that $0\leq k\leq \ell\leq n$ we define $\sigma_k^\ell$ to be $\sigma_k \comp \cdots \comp \sigma_\ell$.
Hence, $\sigma_0^k$ is such that:
$(\Sigma\cup v_\mathcal{X}^{k})^*\xrightarrow{\sigma_{k}}(\Sigma\cup v_\mathcal{X}^{k-1})^*\cdots(\Sigma\cup v_\mathcal{X}^1)^*\xrightarrow{\sigma_1}(\Sigma\cup v_\mathcal{X}^0)^*\xrightarrow{\sigma_0}\Sigma^*$.

Finally, the \emph{$k$-fold composition} of a linear grammar $\tilde{G}$ and
initial variable $X$ is given by
$\sigma_0^k(v_X^k)$. 
%
%
Lemma~\ref{lem-bnusubst} relates $k$-fold compositions with 
$(\bnu_{k})_{k\in\nats}$.
\ifshort
Moreover we characterize the complexity of computing $\tilde{G}$ given a
polynomial transformation $\funbf$ the size of which is defined to be the number
of bits needed to write the set $\set{\funbf_X}_{X\in\mathcal{X}}$ where
each $\funbf_X$ is a string of symbols.
\begin{lemma}
	Given a polynomial transformation $\funbf$, 
	there is a polynomial time algorithm to compute a linear grammar $\tilde{G}$
	such that for every $k\geq 0$, every $X\in\mathcal{X}$ we have
	$\bnu_{k}(X)=\sigma_0^{k}(v_X^k)$.
	\label{lem-bnusubst}
\end{lemma}
\fi
\iflong
\begin{lemma}
	There exists an effectively computable linear grammar $\tilde{G}$ such that
	for every $k\geq 0$, every $X\in\mathcal{X}$ we have
	$\bnu_{k}(X)=\sigma_0^{k}(v_X^k)$.
	\label{lem-bnusubst}
\end{lemma}
\begin{proof}
 By induction on $k$.

\noindent %
{\bf Base case. $(k=0)$} Definition of the iteration sequence shows that
$\bnu_{0}(X)=\funbf(\vzero)(X)$ which in turn equals
$\sigma_0(v_X^0)$ by definition.

\noindent %
{\bf Inductive case. $(k+1)$}
First, let us define $\sigma_{\bnu_k}$ to be the substitution which maps
$v_X$ onto $\bnu_k(X)$. Hence we have
\begin{align*}
	\bnu_{k+1}&=D\deriv{\funbf}_{\bnu_k}^*(\funbf(\bnu_k))&\text{def.\ of $\bnu_{k+1}$}\\
	&=\sigma_{\bnu_k}(L(\tilde{G}))&\text{Lem.~\ref{lem:lingcoincideiterates}, def.\ of $\sigma_{\bnu_k}$}
\intertext{The above definition shows that $\sigma_{\bnu_k}(v_X)=\bnu_k(X)$, hence
that $\sigma_{\bnu_k}(v_X)=\sigma_0^k(v^k_X)$ by induction hypothesis. Hence}
\bnu_{k+1}(X)&=\sigma_{\bnu_k}(L_X(\tilde{G}))\\
&=\sigma_{\bnu_k}\comp\sigma_{[v_Y^k/v_Y]}(\sigma^{k+1}(v_X^{k+1}))&\text{def.\ of $\sigma^{k+1}$}\\
&=\sigma_0^k\comp \sigma^{k+1}(v_X^{k+1})&\text{by above}\\
&=\sigma_0^{k+1}(v_X^{k+1})
\end{align*}
\qed
\end{proof}
\fi

\noindent
\ifshort
We refer the reader to our technical report \cite{CoRR}
for the polynomial time construction of $\tilde{G}$ given $\funbf$. 
However, let us give a sample output of the construction.
\begin{example}
Let $\funbf$ be a polynomial transformation on $\lring^{\mathcal{X}}$
where $\funbf_{X_0}=a X_1 \cup a$ and $\funbf_{X_1}=X_0 b \cup a X_1 b X_0$.
The construction outputs 
$\tilde{G}=(\set{X_0,X_1},\set{a,b,v_{X_0},v_{X_1}},\tilde{\delta})$
where $\tilde{\delta}$ is given by:
\begin{align*}
	&X_0 \rightarrow a X_1 \mid a v_{X_1} \mid a \\
	&X_1 \rightarrow X_0 b \mid a X_1 b v_{X_0} \mid a v_{X_1} b X_0 \mid v_{X_0} b \mid a v_{X_1} b v_{X_0}\enspace .
\end{align*}
We have that $\bnu_1(X_0)=\sigma_0\comp\sigma_1(v_{X_0}^{1})$ and $\bnu_1(X_1)=\sigma_0\comp\sigma_1(v_{X_1}^{1})$.
\end{example}
\fi
Lem.~\ref{lem-bnusubst} completes our goal to define a procedure to effectively
compute and represent the iterates $(\bnu_{k})_{k\in\nats}$.  This sequence is
of interest since, given a CFL $L$ and $\bnu_n$ the $n$-th iterate (where $n$
equals the number of variables in the grammar of $L$ so that
$\Pi(\bnu_n)=\Pi(L)$), if $B$ is a solution to Pb.~\ref{pb:elembounded} for the
instance $\bnu_n$, $B$ is also a solution to Pb.~\ref{pb:elembounded} for $L$.

\iflong
Let us conclude this section on a complexity note.  Below we show that the
linear grammar $\tilde{G}$ given in Lem.~\ref{lem-bnusubst} is computable in
polynomial time in the size of $\funbf$ which is to be defined. To start with
we define the size of a monomial which is intuitively the length of the
``string'' that defines the monomial.
Formally, let $m$ be a monomial its size denoted,
$\sizeof{m}$, is given by $0$ if $m$ is the empty monomial; $1$ if $m\in
2^{\Sigma^*}$ or $m\in\mathcal{X}\cup\mathrm{d}\mathcal{X}$ and by
$\sizeof{m_1}+\sizeof{m_2}$ if $m=m_1\cdot m_2$.  The above definition
naturally extends to polynomials by summing the sizes of the monomials. The
empty polynomial has size zero. 

In what follows we show that the derivative of a monomial as a polynomial of
some form. 
\begin{lemma}
Let $m=b_1 \cdots b_k$ be a monomial where each $b_i\in 2^{\Sigma^*}\cup
\mathcal{X}$, let $X\in\mathcal{X}$ and $\vbf\in \lring^{\mathcal{X}}$.
We have $D_X\deriv{m}_{\vbf}$ coincide with the polynomial given by:
\begin{enumerate}
	\item apply the inductive definition  of a derivative on $m$ which is given
		by $D_X\deriv{m}_{\vbf}= D_X\deriv{(b_1\cdots b_{k-1})}_{\vbf}\cdot
		\vbf(b_{k}) \cup (b_1\cdots b_{k-1}) \cdot a$ where $a=\mathrm{d}X$ if
		$b_k=X$ and $\emptyset$ otherwise. Above we abusively wrote $\vbf(b_k)$
		which in fact denotes $\vbf(b_k)$ if $b_k\in\mathcal{X}$ and $b_k$
		otherwise. 
	\item turn the result into a polynomial, that is a finite combination of
		monomials, by distributing $\cdot$ over $\cup$ (in the inductive part of
		point (1)).
\end{enumerate}
\label{lem-derivmono}
\end{lemma}
In the rest of this section, we identify $D_X\deriv{m}_{\vbf}$ with
the polynomial of Lem.~\ref{lem-derivmono}.

\begin{lemma}
	Let $m=b_1 \cdots  b_k$, and
	$D_X\deriv{m}_{\vbf}=\bigcup_{i\in\set{1,\dots,I}} m_{i}$.  We have
	$\sizeof{m_i}\leq k$ and $I\leq k$.
	\label{lem-sizemono}
\end{lemma}
\begin{proof}
	  \noindent 
	  {\bf $k=1$.}
		$D_X\deriv{m}_{\vbf}=\begin{cases}\mathrm{d}X &\text{if $m=X$}\\ \emptyset
		&\text{else}\end{cases}$ which concludes the case.

		\noindent 
		{\bf $k>1$.} Induction hypothesis shows that $D_X\deriv{(b_1 \cdots
		b_{k-1})}_{\vbf}=\bigcup_{j\in\set{1,\dots,J}} m'_j$ where
		$\sizeof{m'_j}\leq k-1$ and $J\leq k-1$. Hence by Lem.~\ref{lem-derivmono},
		the distributivity of $\cdot$ over $\cup$, the size of $\vbf(b_k)$ bounded
		by 1 show that $\sizeof{m_i}\leq k$ and $I=J+1\leq k$.\qed
\end{proof}

\begin{corollary}
	The size of $D_X\deriv{(b_1 \cdots  b_k)}_{\vbf}$ is bounded by $k^2$ (where
	$k$ is the size of the monomial).
	\label{coro-sizemono}
\end{corollary}

Let us extend this reasoning to polynomials and polynomial transformations.
Let $f=\bigcup_{i\in\set{1,\dots, I}} m_i$.  The definition of differential
shows that $D_X\deriv{f}_{\vbf}=\bigcup_{1\leq i\leq I}
D_X\deriv{m_i}_{\vbf}$ where each $D_X\deriv{m_i}_{\vbf}$ 
 is a polynomial as shown by Lem.~\ref{lem-derivmono}. Let $n=\sizeof{f}$, we
 have that $\sizeof{D_X\deriv{f}_{\vbf}}$ is bounded by $n^3$.  This result
 follows from Coro.~\ref{coro-sizemono} and the fact that $I\leq n$.

Let us now extend our result to the differential in each variable. The
definition of derivative shows that
$D\deriv{f}_{\vbf}=\bigcup_{X\in\mathcal{X}} D_X\deriv{f}_{\vbf}$ the
definition of which is given above.  Let $n=\max(|\mathcal{X}|,\sizeof{f})$, we
find that $\sizeof{D\deriv{f}_{\vbf}}$ is bounded by $n^4$.

Finally we extend the result to polynomial transformation using the equality
$(D\deriv{\funbf}_{\vbf})(X)=D\deriv{\funbf_X}_{\vbf}$.  Let us now
characterize the time complexity of the algorithm that computes for
$D\deriv{\funbf}_{\vbf}$.

\begin{corollary}
	Let $\funbf$ and $\vbf$ be respectively a polynomial transformation and a
	valuation over $\mathcal{X}$. Define $S=\set{\vbf(X)}_{X\in\mathcal{X}}\cup\set{a\in\lring\mid \exists
	X\in\mathcal{X}\colon \text{$a$ occurrs in $\funbf_X$}}$.
	The size of $S$ is given by the sum of the size of each of its member.
	The size of $a\in 2^{\Sigma^*}$ is given by the sum of the length of each
	$w\in a$.  If $S$ is of finite size then
	$D\deriv{\funbf}_{\vbf}$ is computable in time polynomial in the size of each
	$\funbf_X$, $\mathcal{X}$ and the size of $S$.
\end{corollary}

Remark that we could generalize and drop the finiteness requirement for $S$.
For example, regular languages or context-free languages would be admissible
candidates for each element of $S$ because they come with a finite
representation and decision procedure for the tests/operations we need to
compute the differential.  

We showed above how to compute $\tilde{G}$ from $D\deriv{\funbf}_{\vbf}$ and
$\funbf$. So we conclude that $\tilde{G}$ is computable in time polynomial in
the size of each $\funbf_X$, $\mathcal{X}$ and the size of $S$.
\fi
\section{Constructing a Parikh Equivalent Bounded Subset}
\label{sec-construction}

We now show how, given a $k$-fold composition $L'$, to compute an elementary
bounded language $B$ such that $\Pi(L'\cap B)=\Pi(B)$, that is we give an
effective procedure to solve Pb.~\ref{pb:elembounded} for the instance $L'$.
This will complete the solution to Pb.~\ref{pb:elembounded}, hence the proof of
Th.~\ref{theo-main}.  In this section, we give an effective construction of
elementary bounded languages that solve Pb.~\ref{pb:elembounded} first for
regular languages, then for linear languages, and finally for a linear
substitution. 
We start with Lem.~\ref{lem:powerl} %
the proof of which is given \ifshort in \cite{CoRR} but also \fi in \cite{LL79}.
First we need to introduce the notion of semilinear sets.
A set $A\subseteq \nats^n$ is a \emph{linear set} if there exist $c\in
\nats^n$ and $p_1,\ldots,p_k \in \nats^n$ such that $A=\set{c+\sum_{i=1}^k
\lambda_i p_i \mid \lambda_i \in \nats}$: $c$ is called the constant of $A$ and
$p_1,\ldots,p_k$ the periods of $A$.  A \emph{semilinear set} $S$ is a finite
union of linear sets: $S=\bigcup_{j=1}^{\ell} A_j$ where each $A_j$ is a linear set.
Parikh's theorem (cf. \cite{ginsburg}) shows that the Parikh image of every CFL
is a semilinear set that is effectively computable.

\begin{lemma}{} 
Let $L$ and $B$ be respectively a CFL and an elementary bounded language over
$\Sigma$ such that $\Pi(L\cap B)=\Pi(L)$.  There is an effectively computable
elementary bounded language $B'$ such that $\Pi(L^t\cap B')=\Pi(L^t)$ for all
 $t\in\nats$.
\label{lem:powerl}
\end{lemma}
\begin{proof}
	By Parikh's theorem, we know that $\Pi_{\Sigma}(L)$ is
a computable semilinear set.
\begin{comment}
\ie a finite union $\bigcup_{i=1}^\ell A_i$
of sets of the form $A_i=\set{c_i+\sum_{j=1}^{k_i} \lambda_j p_{ij} \mid \lambda_j \in \nats}$
with $c_i\in \nats^{|\Sigma|}$ the constant of
$A_i$ and $p_{i1},\ldots,p_{ik_i}\in \nats^{|\Sigma|}$ the periods
of $A_i$.
\end{comment}
Let us consider $u_1,\ldots,u_\ell \in L$ such
that $\Pi_{\Sigma}(u_i)=c_i$ for $i\in \set{1,\ldots,\ell}$.

Let $B'=u_1^*\cdots u_\ell^* B^\ell$, we see that $B'$ is an elementary bounded
language. Let $t>0$ be a natural integer. We have to prove that $\Pi(L^t)\subseteq \Pi(L^t\cap B')$.

\noindent $\boldsymbol{t\leq \ell}$ 
\iflong
We conclude from the preservation of $\Pi$
and the hypothesis $\Pi(L)=\Pi(L\cap B)$ that
\fi
\ifshort
By property of $\Pi$ and $\Pi(L)=\Pi(L\cap B)$ we find that:
\fi 
\begin{align*}
	\Pi(L^t)&= \Pi((L\cap B)^t)\\
	&\subseteq \Pi(L^t\cap B^t)&\text{monotonicity of $\Pi$}\\
	&\subseteq \Pi(L^t\cap B^\ell)&\text{$B^t\subseteq B^\ell$ since $\varepsilon \in B$}\\
	&\subseteq \Pi(L^t\cap B')&\text{def.\ of $B'$}
\end{align*}
$\boldsymbol{t>\ell}$ Let us consider $w\in L^t$. For every $i\in\set{1,\ldots,\ell}$
and $j\in\set{1,\ldots,k_i}$, there exist some positive
integers $\lambda_{ij}$ and $\mu_{i}$, with $\sum_{i=1}^\ell \mu_i = t$ such that
\[\Pi(w)=\sum_{i=1}^\ell \mu_i c_i + \sum_{i=1}^\ell \sum_{j=1}^{k_i} \lambda_{ij} p_{ij}\enspace .\]
We define a new variable for each $i\in\set{1,\ldots,\ell}$:
$\alpha_i =
\begin{cases}
  \mu_i -1  &\text{ if } \mu_i > 0 \\
  0 	    &\text { otherwise.}
\end{cases}$.\\
For each $i\in\set{1,\ldots,\ell}$, we also consider $z_i$ a word of
$L\cup\set{\varepsilon}$ such that $z_i=\varepsilon$ if $\mu_i=0$ and
$\Pi(z_i)=c_i+\sum_{j=1}^{k_i} \lambda_{ij} p_{ij}$ else.

Let $w'=u_1^{\alpha_1}\ldots u_\ell^{\alpha_\ell} z_1\ldots z_\ell$.
Clearly, $\Pi(w')=\Pi(w)$ and $w'\in u_1^*\cdots u_\ell^* (L\cup\set{\varepsilon})^\ell$.
For each $i\in \set{1,\ldots,\ell}$,
$\Pi(L\cap B)=\Pi(L)$ shows that there is $z'_i\in(L\cap B)\cup\set{\varepsilon}$
such that $\Pi(z'_i)=\Pi(z_i)$.
Let $w''=u_1^{\alpha_1}\ldots u_\ell^{\alpha_\ell} z'_1\ldots z'_\ell$.
We find that $\Pi(w'')=\Pi(w)$, $w''\in B'$ and we can easily verify that $w''\in L^t$.
\qed
\end{proof}

\smallskip%
\noindent%
{\bf Regular Languages.}%
The construction of an elementary bounded language that solves
Pb.~\ref{pb:elembounded} for a regular language $L$ is known from \cite{LL79}
(see also \cite{ls04}, Lem.~4.1). The construction is carried out by induction
on the structure of a regular expression for $L$. 
Assuming $L\neq\emptyset$, the base case (\ie a symbol or $\varepsilon$) is
trivially solved. Note that if $L=\emptyset$ then every elementary bounded language $B$ is such that $\Pi(L\cap B)=\Pi(L)=\emptyset$.

The inductive case falls naturally into three parts.
Let $R_1$ and $R_2$ be regular languages, and $B_1$ and $B_2$ the
inductively constructed elementary bounded languages
such that $\Pi(R_1\cap B_1)=\Pi(R_1)$ and $\Pi(R_2\cap B_2)=\Pi(R_2)$.
\begin{description}
\item[concatenation] For the
  instance $R_1\cdot R_2$, the elementary bounded
  language $B_1\cdot B_2$ is such that
  $\Pi( (R_1\cdot R_2)\cap (B_1\cdot B_2))=\Pi( R_1\cdot R_2)$;
\item[union] For $R_1 \cup R_2$, the elementary bounded language $B_1\cdot
  B_2$ suffices;
\item[Kleene star] Let us consider $R_1$ and $B_1$,
  Lem.~\ref{lem:powerl} shows how to effectively compute
  an elementary bounded language $B'$ such that for every
  $t\in\nats$, $\Pi(R_1^t \cap B')=\Pi(R_1^t)$.
  Let us prove that $B'$ solves
  Pb.~\ref{pb:elembounded} for the instance $R_1^*$. In
  fact, if $w$ is a word of $R_1^*$, there exists a
  $t\in\nats$ such that $w\in R_1^t$. Then, we can find a
  word $w'$ in $R_1^t\cap B'$ with
  the same Parikh image as $w$. This proves that
  $\Pi(R_1^*)\subseteq \Pi(R_1^*\cap B')$. The other
  inclusion holds trivially.
\end{description}

\begin{proposition}
For every regular language $R$,
there is an effective procedure to compute an
elementary bounded language $B$ such that $\Pi(R\cap B)=\Pi(R)$.
\label{prop:regular}
\end{proposition}

\smallskip%
\noindent%
{\bf Linear Languages.}%
We now extend the previous construction to the case of linear languages.
Recall that linear languages are used to represent the iterates
$(\bnu_{k})_{k\in\nats}$. Lemma~\ref{prop:linear} gives a characterization of
linear languages based on regular languages, homomorphism, and some
additional structures.

\begin{lemma}{(from \cite{HU79})}
For every linear language $L$ over $\Sigma$, there exist an alphabet $A$ and
its distinct copy $\widetilde{A}$, an homomorphism $h:(A\cup \widetilde{A})^*
\rightarrow \Sigma^*$ and a regular language $R$ over $A$ such that
$L=h(R\widetilde{A}^* \cap S)$ where $S=\{w \widetilde{w}^r \mid w \in A^*
\}$ and $w^r$ denotes the reverse image of the word $w$.
Moreover there is an effective procedure to construct $h$, $A$, and $R$.
\label{prop:linear}
\end{lemma}
\iflong
\begin{proof}
	Assume the linear language $L$ is given by linear grammar $G=(\mathcal{X},\Sigma,\delta)$ and a initial variable $X_0$.
	We define the alphabet $A$ to be $\set{ a_p\mid p\in\delta}$.
	We define the regular language $R$ as the language accepted by
	the automaton given by $(\mathcal{X}\cup\set{q_f}, T, X_0, \set{q_f})$
	where: $T=\set{(X,a_p,Y)\mid p=(X,\alpha Y\beta)\in\delta}\cup\set{(X,a_p,q_f)\mid p=(X,\alpha)\in\delta \land \alpha\in\Sigma^*}$.
	Next we define the homomorphism, $h$ which, for each $p=(X,\alpha
	Y\beta)\in\delta$, maps $a_p$ and $\widetilde{a_p}$ to $\alpha$ and
	$\beta$, respectively.
	By construction and induction on the length of a derivation, it is easily seen that the result holds.
\qed
\end{proof}

Next, we have a technical lemma which relates homomorphism and the Parikh
image operator.
\begin{lemma}
	Let $X,Y\subseteq\Sigma^*$ be two languages and 	
a homomorphism $h:A^* \rightarrow \Sigma^*$, we have:
\[\Pi(X)=\Pi(Y)\text{ implies } \Pi(h(X))=\Pi(h(Y))\enspace .\]
	\label{lem:homomorph}
\end{lemma}
\begin{proof}
	It suffices to show that the result holds for $=$ replaced
	by $\subseteq$.
	Let $x'\in h(X)$. We know that there exists $x\in X$ such that
	$x'=h(x)$. The equality $\Pi(X)=\Pi(Y)$ shows that there exists
	$y\in Y$ such that $\Pi(y)=\Pi(x)$. 	
	It is clear by property of homomorphism that $\Pi(h(y))=\Pi(h(x))$.\qed
\end{proof}

\fi

The next result shows that an elementary bounded language that solves
Pb.~\ref{pb:elembounded} can be effectively constructed for every linear
language $L$ that is given by $h$ and $R$ such that $L=h(R\widetilde{A}^* \cap
S)$.
\begin{proposition}
For every linear language
$L=h(R\widetilde{A}^* \cap S)$ where $h$ and $R$ are given, there is an
effective procedure which solves Pb.~\ref{pb:elembounded} for the instance
$L$, that is a procedure returning an elementary bounded $B$ such that
$\Pi(L\cap B)=\Pi(L)$.
\label{prop:linbounded}
\end{proposition}
\iflong
\begin{proof}
Since $R$ is a regular language, we can use the result of
Prop.~\ref{prop:regular} to effectively compute the set $\set{w_1,\ldots,w_m}$
of words such that for $R'=R\cap w_1^*\cdots w_m^*$ we have $\Pi(R')=\Pi(R)$.
Also, we observe that for every language $Z\subseteq A^*$ we have
$Z\widetilde{A}^*\cap S=\set{w\widetilde{w}^r\mid w\in Z}$.
\begin{align*} &\Pi(R')=\Pi(R) &\text{by above}\\
&\text{only if }\Pi(R'\widetilde{A}^*\cap S)=\Pi(R\widetilde{A}^*\cap S)&\text{by above}\\
&\text{only if }\Pi(h(R'\widetilde{A}^*\cap S))=\Pi(h(R\widetilde{A}^*\cap S))&
\text{Lem.~\ref{lem:homomorph}}\\
&\text{only if }\Pi(h(R'\widetilde{A}^*\cap S))=\Pi(L)&\text{def.\ of $L$}\\
&\text{only if }\Pi(h(R\widetilde{A}^*\cap S)\cap w_1^*\cdots w_m^* {\widetilde{w_m^r}}^*\cdots {\widetilde{w_1^r}}^*)=\Pi(L)&\text{def. of $R'$}\\
&\text{only if }\Pi(h(R\widetilde{A}^*\cap S)\cap h(w_1^*\cdots w_m^* {\widetilde{w_m^r}}^*\cdots {\widetilde{w_1^r}}^*))=\Pi(L)\\
&\text{only if }\Pi(L\cap h(w_1^*\cdots w_m^* {\widetilde{w_m^r}}^*\cdots {\widetilde{w_1^r}}^*))=\Pi(L)&\text{def.\ of $L$}\\
&\text{only if }\Pi(L\cap h(w_1)^*\cdots h(w_m)^* {h(\widetilde{w_m^r}})^*\cdots h({\widetilde{w_1^r}})^*)=\Pi(L)
\end{align*}
which concludes the proof since $h(w)\in\Sigma^*$ if $w\in
(A\cup\widetilde{A})^*$.
\qed
\end{proof}
\fi

\smallskip%
\noindent%
{\bf Linear languages with Substitutions.}%
Our goal is to solve Pb.~\ref{pb:elembounded} for $k$-fold compositions, \ie
for languages of the form $\sigma_j^k(v_X^k)$. Prop.~\ref{prop:linbounded}
gives an effective procedure for the case $j=k$ since $\sigma_k^k(v_X^k)$ is a
linear language.
Prop.~\ref{prop:subst} generalizes to the case
$j<k$: given a solution to Pb.~\ref{pb:elembounded} for the instance
$\sigma_{j+1}^k(v_X^k)$, there is an effective procedure for
Pb.~\ref{pb:elembounded} for the instance $\sigma_j\comp\sigma_{j+1}^k(v_X^k)=\sigma_{j}^k(v_X^k)$.

\begin{proposition}
	Let
\begin{enumerate}
	\item $L$ be a CFL over $\Sigma$;
	\item $B$ an elementary bounded language such that $\Pi(L\cap B)=\Pi(L)$;
	\item $\sigma$ and $\tau$ be two substitutions over $\Sigma$
such that for each $a\in\Sigma$, (i) $\sigma(a)$ and $\tau(a)$ are
respectively a CFL and an e.b. and
(ii) $\Pi(\sigma(a)\cap\tau(a))=\Pi(\sigma(a))$.
\end{enumerate}
Then, there is an effective procedure that solves Pb.~\ref{pb:elembounded}
for the instance $\sigma(L)$, by returning an elementary
bounded language $B'$ such that $\Pi(\sigma(L)\cap B')=\Pi(\sigma(L))$.
\label{prop:subst}
\end{proposition}
\iflong
\begin{proof}
Let $w_1,\ldots,w_k \in \Sigma^*$ be the words such that $B=w_1^*\cdots w_k^*$.
Let $L_i=\sigma(w_i)$ for each $i\in\set{1,\ldots,k}$.  Since $\sigma(a)$ is a
CFL so is $\sigma(w_i)$ by property of the substitutions and
the closure of CFLs by finite concatenations. For the same
reason, $\tau(w_i)$ is an elementary bounded language. Next, Lem.~\ref{lem:powerl}
where the elementary bounded language is given by $\tau(w_i)$,
shows that we can construct an elementary bounded language $B_i$ such that for all
$t\in\nats$, $\Pi(L_i^t\cap B_i)=\Pi(L_i^t)$. Define $B'=B_1\ldots
B_k$ that is an elementary bounded language.
We have to prove the inclusion $\Pi(\sigma(L))\subseteq \Pi(\sigma(L)\cap B')$ since the reverse one trivially holds.
So, let $w\in \sigma(L)$. Since $\Pi(L\cap w_1^*\cdots w_k^*)=\Pi(L)$, there
is a word $w'\in\sigma(L\cap w_1^*\cdots w_k^*)$ such that $\Pi(w)=\Pi(w')$.
Then we have
\begin{align*}
w'&\in \sigma(L\cap w_1^*\cdots w_k^*)\\
&\in \sigma(w_1^{t_1}\dots w_k^{t_k})&\text{for some $t_1,\dots,t_k$}\\
&\in \sigma(w_1^{t_1})\dots \sigma(w_k^{t_k}) &\text{property of subst.}\\
&\in \sigma(w_1)^{t_1}\dots \sigma(w_k)^{t_k} &\text{property of subst.}\\
&\in L_1^{t_1}\dots L_k^{t_k} &\text{$\sigma(w_i)=L_i$}
\end{align*}
For each $i\in\set{1,\ldots,k}$, we have
$\Pi(L_i^{t_i}\cap B_i)=\Pi(L_i^{t_i})$, so we can find $w''\in (L_1^{t_1}\cap
B_1)\ldots (L_k^{t_k}\cap B_k)$ such that $\Pi(w'')=\Pi(w')$.
Definition of $B'$ also shows that $w''\in B'$.
Moreover
\begin{align*}
w''&\in (L_1^{t_1}\cap B_1)\ldots (L_k^{t_k}\cap B_k)\\
&\in L_1^{t_1}\ldots L_k^{t_k}\\
&\in \sigma(w_1)^{t_1}\ldots \sigma(w_k)^{t_k}&\text{$\sigma(w_i)=L_i$}\\
&\in \sigma(w_1^{t_1})\ldots \sigma(w_k^{t_k})&\text{property of subst.}\\
&\in \sigma(w_1^{t_1}\ldots w_k^{t_k})&\text{property of subst.}\\
&\in \sigma(L\cap w_1^{*}\ldots w_k^{*})&\text{$w_1^{t_1}\ldots w_k^{t_k}\in L\cap w_1^{*}\ldots w_k^{*}$}\\
&\in \sigma(L)
\end{align*}
Finally, $w''\in B'$ and $w''\in\sigma(L)$ and $\Pi(w'')=\Pi(w')$, which
in turn equals $\Pi(w)$, prove the inclusion.
\qed
\end{proof}
\fi

We use the above result inductively to solve Pb.~\ref{pb:elembounded} for
$k$-fold composition as follows: fix $L$ to be $\sigma_{j+1}^k(v_X^k)$, $B$ to
be the solution of Pb.~\ref{pb:elembounded} for the instance $L$, $\sigma$ to
be $\sigma_j$ and $\tau$ a substitution which maps every $v_X^j$ to the
solution of Pb.~\ref{pb:elembounded} for the instance $\sigma_j(v_X^j)$. Then
$B'$ is the solution of Pb.~\ref{pb:elembounded} for the instance
$\sigma_j^k(v_X^k)$. %

\iflong
\subsection{$k$-fold Substitutions}

Let us now solve Pb.~\ref{pb:elembounded} where the instance is given by a
$k$-fold composition.  Given a CFL $L=L_{X_0}(G)$ where
$G=(\mathcal{X},\Sigma,\delta)$ is a grammar and $X_0\in \mathcal{X}$ an
initial variable, we compute the linear grammar $\tilde{G}$ and the $k$-fold
composition $\set{\sigma_j}_{0\leq j\leq n}$ as defined in
Sec.~\ref{subsec:representation}. With the result of
Prop.~\ref{prop:linbounded}, we find a valuation $\tilde{B}$ such that for
every variable $X$, (1) $\tilde{B}(X)$ is an elementary bounded language and
(2) $\Pi(L_X(\tilde{G}))=\Pi(L_X(\tilde{G})\cap \tilde{B}(X))$.

The above reasoning is formally explained in Alg.~\ref{alg:boundedsequence}.

\begin{algorithm}[t]
	\caption{Bounded Sequence\label{alg:boundedsequence}}
  \KwData{$\tilde{G}$ a linear grammar}
  \KwData{$\tilde{B}$ a valuation s.t. for every $X\in \mathcal{X}$ $\tilde{B}(X)$ is an elementary bounded language and $\Pi(L_X(\tilde{G}))=\Pi(L_X(\tilde{G})\cap \tilde{B}(X))$}
  \KwData{$n\in\nats$}
  \KwResult{$B\in\lring^{\mathcal{X}}$  such that for every $X\in\mathcal{X}$ $B(X)$ is an elementary bounded and $\Pi(B(X)\cap \bnu_n(X))=\Pi(\bnu_n(X))$ }
  \lnl{alg:Bnmoins1}Let $B_{n-1}$ be $\tilde{B}[v_\mathcal{X}^{n-1}]$\;
  \For{$i=n-2,n-3,\dots,0$}{
  Let $\tau_{i+1}$ be the substitution which maps each $v_X^{i+1}$ on $\tilde{B}[v_\mathcal{X}^i]$ and leaves each letter of $\Sigma$ unchanged\;
    \ForEach{$X\in \mathcal{X}$}{
       \lnl{alg:biplus1} Let $B_i(X)$ be the language returned by Prop.~\ref{prop:subst} on the languages
       $\sigma_{i+2}^{n}(v_X^n)$ and $B_{i+1}(X)$, and
       the substitutions $\sigma_{i+1},\tau_{i+1}$\;
    }
  }
  Let $\tau_0$ be the substitution which maps each $v_X^0$ on the elementary bounded language $w_1^*\cdots w_p^*$ where $\set{w_1,\dots,w_p}=\sigma_0(v_X^0)$ and leaves each letter of $\Sigma$ unchanged\;
  \ForEach{$X\in \mathcal{X}$}{
  \lnl{alg:bx} Let $B(X)$ be the language returned by Prop.~\ref{prop:subst}
  on the languages $\sigma_1^n(v_X^n)$ and $B_0(X)$, and the substitutions
  $\sigma_0$, $\tau_0$\;
  }
  \KwRet{B}
\end{algorithm}

We now prove the following invariants for Alg.~\ref{alg:boundedsequence}.

\begin{lemma}
	In Alg.~\ref{alg:boundedsequence}, for every $X\in\mathcal{X}$,
	\begin{itemize}
	\item for every $k\in\set{0,\ldots,n-1}$,
		$B_k(X)$ is an elementary bounded language on $(\Sigma\cup v_\mathcal{X}^k)^*$ such that
	 $\Pi(\sigma_{k+1}^{n}(v_X^n)\cap B_{k}(X))=\Pi(\sigma_{k+1}^{n}(v_X^n))$;

	\item $B(X)$ is an elementary bounded language on $\Sigma^*$ such that $\Pi(\bnu_n(X)\cap B(X))=\Pi(\bnu_n(X))$.
	\end{itemize}
\end{lemma}
\begin{proof}
\begin{itemize}
\item By induction on $k$:

\noindent {\bf Base case. $(k=n-1)$} Alg.~\ref{alg:boundedsequence} assumes that $\tilde{B}(X)$ is an elementary bounded language, so is $B_{n-1}$ by line~\ref{alg:Bnmoins1}.
It remains to prove that $\Pi(\sigma_{n}(v_X^n)\cap B_{n-1}(X))=\Pi(\sigma_{n}(v_X^n))$, which is equivalent, by definition of $\sigma_n$ and $B_{n-1}$, to
$\Pi(L_X(\tilde{G})[v_\mathcal{X}^{n-1}]\cap \tilde{B}[v_\mathcal{X}^{n-1}](X))=\Pi(L_X(\tilde{G})[v_\mathcal{X}^{n-1}])$.
By property of the symbol-to-symbol substitution $\sigma_{[v_Y/v_Y^{n-1}]}$, the equality reduces to
$\Pi(L_X(\tilde{G})\cap \tilde{B}(X))=\Pi(L_X(\tilde{G}))$ which holds by assumption of Alg.~\ref{alg:boundedsequence}.

\noindent {\bf Inductive case. $(0\leq k \leq n-2)$} At line~\ref{alg:biplus1}, we see that we can apply the result of Prop.~\ref{prop:subst} because (1) $\sigma_{i+2}^{n}(v_X^n)$ is a CFL (CFLs are closed by context-free substitutions), (2) $B_{i+1}(X)$ is an elementary bounded language (induction hypothesis), (3) for every variable $Y\in\mathcal{X}$, $\sigma_{i+1}(v_Y^{i+1})$ is a CFL, $\tau_{i+1}(v_Y^{i+1})$ is an elementary bounded language and $\Pi(\sigma_{i+1}(v_Y^{i+1})\cap
\tau_{i+1}(v_Y^{i+1}))= \Pi(\sigma_{i+1}(v_Y^{i+1}))$. Hence, the proposition shows that $B_i(X)$ is an
elementary bounded language and
$\Pi(\sigma_{i+1}^{n}(v_X^n)\cap B_{i}(X))=\Pi(\sigma_{i+1}^{n}(v_X^n))$.

\item The above invariant for $k=0$ shows that, for every variable $X\in\mathcal{X}$,
(1) $B_0(X)$ is an elementary bounded language,
and (2) $\Pi(\sigma_1^n(v_X^n)\cap B_0(X))=\Pi(\sigma_1^n(v_X^n))$.
We conclude from line~\ref{alg:bx} and Prop.~\ref{prop:subst} that
$\Pi(\sigma_0^n(v_X^n)\cap B(X))=\Pi(\sigma_0^n(v_X^n))$, and that $\Pi(\bnu_n(X)\cap B(X))=\Pi(\bnu_n(X))$
by Lem.~\ref{lem-bnusubst}.
\end{itemize}
\qed
\end{proof}

Referring to our initial problem, we finally find that:
\begin{corollary}
Let $B$ be the valuation returned by Alg.~\ref{alg:boundedsequence},
$B$ is a valuation in $\lring^{\mathcal{X}}$ such that for every $X\in
\mathcal{X}\colon \Pi(L_X(G)\cap B(X))=\Pi(L_X(G))$.
\end{corollary}

In fact, for $X=X_0$, $B(X_0)$ is the solution of Pb.~\ref{pb:elembounded} for the instance $L$.
This concludes the proof of Th.~\ref{theo-main}.
In what follows, we show two applications of Th.~\ref{theo-main}
in software verification.
\fi
\ifshort
Due to lack of space we refer to reader to \cite{CoRR} for details.

We thus have an effective construction of an elementary bounded language that
solves Pb.~\ref{pb:elembounded} for $k$-fold composition, hence a constructive
proof for Th.~\ref{theo-main}.%
\fi

\smallskip
\noindent
{\bf Iterative Algorithm.}
We conclude this section by showing a result
related to the notion of progress if the result of Th.~\ref{theo-main} is
applied repeatedly.
\begin{lemma}
	Given a CFL $L$, define two sequences
	$(L_i)_{i\in\mathbb{N}}$, $(B_i)_{i\in\mathbb{N}}$ such that (1) $L_0=L$, (2)
	$B_i$ is elementary bounded and $\Pi(L_i\cap B_i)=\Pi(L_i)$, (3)
	$L_{i+1}=L_i\cap \overline{B_i}$.  For every $w\in L$, there exists
	$i\in\mathbb{N}$ such that $w\notin L_{i}$.
	Moreover, given $L_0$, there is an effective procedure to compute $L_i$ for
	every $i>0$.
	\label{lem:progress}
\end{lemma}
%
\begin{proof}
	Let $w\in L$ and let $v=\Pi(w)$ be its Parikh image.  We conclude form
	$\Pi(L_0\cap B_0)=\Pi(L_0)$ that there exists a word $w'\in B_0$ such that
	$\Pi(w')=v$. Two cases arise: either $w'=w$ and we are done; or $w'\neq w$.
	In that case $L_1=L_0\cap\overline{B_0}$ shows that $w'\notin L_1$.  Intuitively, at
	least one word with the same Parikh image as $w$ has been selected by $B_0$
	and then removed from $L_0$ by definition of $L_1$.  Repeatedly applying the
	above reasoning shows that at each iteration there exists a word $w''$ such
	that $\Pi(w'')=v$, $w''\in B_i$ and $w''\notin L_{i+1}$ since
	$L_{i+1}=L_i\cap\overline{B_i}$. Because there are only finitely many words
	with Parikh image $v$ we conclude that there exists $j\in\nats$, such that
	$w\notin L_{j}$.  The effectiveness result follows from the following
	arguments: (1) as we have shown above (our solution to
	Pb.~\ref{pb:elembounded}), given a CFL $L$ there is an effective procedure
	that computes an elementary bounded language $B$ such that $\Pi(L\cap
	B)=\Pi(L)$; (2) the complement of $B$ is a regular language effectively
	computable; and (3) the intersection of a CFL with a regular language is
	again a CFL that can be effectively constructed (see \cite{HU79}). 
	\qed
\end{proof}
%
Intuitively this result shows that given a context-free language $L$, if we
repeatedly compute and remove a Parikh-equivalent bounded subset of $L$
($L\cap \overline{B}$ is effectively computable since $B$ is a regular
language), then each word $w$ of $L$ is eventually removed from it.

\iflong
\section{Applications}\label{sec-applications}

We now demonstrate two applications of our construction.  The first application
gives a semi-algorithm for checking reachability of multithreaded procedural
programs \cite{Ramalingam,Kahlon,BET03}.  The second application computes an
underapproximation of the reachable states of a recursive counter machine. 
\fi

\ifshort 
\section{Application to Multithreaded Procedural Programs}\label{sec-applications}
We now give an application of our construction that gives a
semi-algorithm for checking reachability of multithreaded procedural programs
\cite{Ramalingam,Kahlon,BET03}.
\fi
\iflong
\subsection{Multithreaded Procedural Programs}

\noindent{\bf Multithreaded Reachability.}
\fi
A common programming model consists of multiple recursive
threads communicating via shared memory.
Formally, we model such systems as pushdown networks \cite{SES08}.
Let $n$ be a positive integer, a \emph{pushdown network} is a triple
$\mathcal{N}=(G,\Gamma,(\Delta_i)_{1\leq i\leq n})$ where $G$ is a finite non-empty set of
\emph{globals}, $\Gamma$ is the \emph{stack alphabet}, and for each $1\leq i\leq n$,
$\Delta_i$ is a finite set of \emph{transition rules} of
the form $\tuple{g,\gamma}\hookrightarrow \tuple{g',\alpha}$ for $g,g'\in G$,
$\gamma\in \Gamma$, $\alpha\in\Gamma^*$.

A \emph{local configuration} of $\mathcal{N}$ is a pair $(g,\alpha)\in G\times
\Gamma^*$ and a \emph{global configuration} of $\mathcal{N}$ is a tuple
$(g,\alpha_1,\dots,\alpha_n)$, where $g\in G$ and
$\alpha_1,\dots,\alpha_n\in\Gamma^*$ are individual stack content for each
thread.  Intuitively, the system consists of $n$ threads, each of which have
its own stack, and the threads can communicate by reading and manipulating the
global storage represented by $g$.

We define the local transition
relation of the $i$-th thread, written $\rightarrow_i$, as follows:
$(g,\gamma\beta)\rightarrow_i (g',\alpha\beta)$ if{}f
$\tuple{g,\gamma}\hookrightarrow\tuple{g',\alpha}$ in $\Delta_i$ and
$\beta\in\Gamma^*$. The transition relation of $\mathcal{N}$, denoted
$\rightarrow$, is defined as follows:
$(g,\alpha_1,\dots,\alpha_i,\dots,\alpha_n)\rightarrow(g',\alpha_1,\dots,\alpha'_i,\dots,\alpha_n)$
if{}f $(g,\alpha_i)\rightarrow_i(g',\alpha'_i)$.  By $\rightarrow_i^*$,
$\rightarrow^*$, we denote the reflexive and transitive closure of these
relations.
Moreover, we define the global reachability relation $\leadsto$ as a
reachability relation where all the moves are made by a single thread:
$(g,\alpha_1,\dots,\alpha_i,\dots,\alpha_n)\leadsto
(g',\alpha_1,\dots,\alpha'_i,\dots,\alpha_n)$ if{}f
$(g,\alpha_i)\rightarrow_i^*(g',\alpha'_i)$ for some $1\leq i\leq n$. The relation $\leadsto$ holds between global configurations reachable from each other in a single
\emph{context}.
Furthermore we denote by $\leadsto_j$, where $j\geq 0$, the reachability
relation within $j$ contexts: $\leadsto_0$ is the identity relation on global
configurations, and $\leadsto_{i+1}=\mathop{\leadsto_{i}}\comp \mathop{\leadsto}$.  Let $C_0$ and $C$ be two global
configurations, the \emph{reachability problem} asks whether $C_0\rightarrow^*
C$ holds. An instance of the reachability problem is denoted by a triple
$(\mathcal{N},C_0,C)$.

A \emph{pushdown system} is a pushdown network where $n=1$, namely
$(G,\Gamma,\Delta)$.  A \emph{pushdown acceptor} is a pushdown system extended
with an initial configuration $c_0\in G\times \Gamma^*$, labeled transition
rules of the form
$\tuple{g,\gamma}\stackrel{\lambda}{\hookrightarrow}\tuple{g'\alpha}$ for
$g,g',\gamma,\alpha$ defined as above and $\lambda\in
\Sigma\cup\set{\varepsilon}$. A pushdown
acceptor is given by a tuple $(G,\Gamma,\Sigma,\Delta,c_0)$. The language of a
pushdown acceptor is defined as expected where the acceptance condition is
given by the empty stack.

In what follows, we reduce the reachability problem for a pushdown network of
$n$ threads to a language problem for $n$ pushdown acceptors.  The pushdown
acceptors obtained by reduction from the pushdown network settings have a
special global $\bot$ that intuitively models an inactive state.  The reduction
also turns the globals into input symbols which label transitions. The firing
of a transition labeled with a global models a context switch.  When such
transition fires, every pushdown acceptor synchronizes on the label.  The
effect of such a synchronization is that exactly one acceptor will change its
state from inactive to active by updating the value of its global (i.e. from
$\bot$ to some $g\in G$) and exactly one acceptor will change from active to
inactive by updating its global from some $g$ to $\bot$. All the others
acceptors will synchronize and stay inactive.

Given an instance of the reachability problem, that is a pushdown network
$(G,\Gamma,(\Delta_i)_{1\leq i\leq n})$ with $n$ threads, two global configurations
$C_0$ and $C$ (assume wlog that $C$ is of the form
$(g,\varepsilon,\dots,\varepsilon)$), we define a family of pushdown acceptors
$\set{(G',\Gamma,\Sigma,\Delta'_i,c^i_0)}_{1\leq i \leq n}$, where:
\begin{itemize}
	\item $G'=G\cup\set{\bot}$, $\Gamma$ is given as above, and $\Sigma=G\times\set{1,\dots,n}$,
	\item $\Delta'_i$ is the smallest set such that:
		\begin{itemize}
			\item
				$\tuple{g,\gamma}\stackrel{\varepsilon}{\hookrightarrow}\tuple{g',\alpha}$
				in $\Delta'_i$ if  $\tuple{g,\gamma}\hookrightarrow\tuple{g',\alpha}$
				in $\Delta_i$;
			\item $\tuple{g,\gamma}\stackrel{(g,j)}{\hookrightarrow}\tuple{\bot,\gamma}$ for $j\in\set{1,\dots,n}\setminus\set{i}$, $g\in G$, $\gamma\in\Gamma$;
			\item $\tuple{\bot,\gamma}\stackrel{(g,j)}{\hookrightarrow}\tuple{\bot,\gamma}$ for $j\in\set{1,\dots,n}\setminus\set{i}$, $g\in G$, $\gamma\in\Gamma$;
			\item $\tuple{\bot,\gamma}\stackrel{(g,i)}{\hookrightarrow}\tuple{g,\gamma}$ for $g\in G$, $\gamma\in\Gamma$.
		\end{itemize}
	\item let $C_0=(g,\alpha_1,\dots,\alpha_i,\dots,\alpha_n)$, $c^i_0$ is given by $(\bot,\alpha_i)$ if $i>1$; $(g,\alpha_1)$ else.
\end{itemize}

\begin{proposition}
Let $n$ be a positive integer, and $(\mathcal{N},C_0,C)$ be an instance of
the reachability problem with $n$ threads, one can
effectively construct CFLs $(L_1,\ldots,L_n)$ (as pushdown acceptors) such
that $C_0\rightarrow^* C$ if{}f $L_1\cap \dots\cap L_n\neq\emptyset$.
\label{prop:cflencoding}
\end{proposition}

The converse of the proposition is also true, and since the emptiness problem
for intersection of CFLs is undecidable \cite{HU79}, so is the
reachability problem.  We
will now compare two underapproximation techniques. The context-bounded
switches for the reachability problem \cite{QR} and the bounded languages for
the emptiness problem that is given below.

Let $L_1,\ldots,L_k$ be context-free languages, and consider the problem to
decide if $\bigcap_{1\leq i \leq k} L_i \neq\emptyset$.
We give a decidable sufficient condition: given an elementary bounded language
$B$, we define the \emph{intersection modulo $B$} of the languages
$\set{L_i}_i$ as $\bigcap_i^{(B)} L_i = \bigl(\bigcap_i L_i \bigr) \cap B$.
Clearly, $\bigcap_i^{(B)} L_i\neq \emptyset$ implies $\bigcap_i L_i
\neq\emptyset$.  Below we show that the problem $\bigcap_i^{(B)} L_i\neq\emptyset$
is decidable .

\begin{lemma}
Given an elementary bounded language $B = w_1^*\cdots w_n^*$ and
CFLs $L_1,\ldots,L_k$,
it is decidable to check if $\bigcap_{1\leq i\leq k}^{(B)} L_i\neq\emptyset$.
\label{lem:capcfleb}
\end{lemma}
\begin{proof}
Define the alphabet $A = \set{a_1,\ldots,a_n}$ disjoint from $\Sigma$.
Let $h$ be the homomorphism that maps the symbols
$a_1,\ldots,a_n$ to the words $w_1,\ldots,w_n$, respectively.
We show that
$\bigcap_{1\leq i\leq k} \Pi_A\bigl(h^{-1}(L_i\cap B)\cap a_1^*\cdots a_n^*\bigr) \neq\emptyset$
if{}f $\bigcap_{1\leq i\leq k}^{(B)} L_i\neq\emptyset$.

We conclude from $w\in \bigcap_{1\leq i\leq k}^{(B)} L_i$ that $w\in B$ and
$w\in L_i$ for every $1\leq i\leq k$, hence there exist $t_1,\ldots,t_n \in
\nats$ such that $w=w_1^{t_1} \ldots w_n^{t_n}$ by definition of $B$.  Then, we find that
$(t_1,\ldots,t_n)=\Pi_A(h^{-1}(w)\cap a_1^*\cdots a_n^*)$, hence that $(t_1,\ldots,t_n)\in
\Pi_A(h^{-1}(L_i\cap B)\cap a_1^*\cdots a_n^*)$ for every $1\leq i\leq k$  by above and finally that
$\bigcap_{1\leq i\leq k} \Pi_A\bigl(h^{-1}(L_i\cap B)\cap a_1^*\cdots
a_n^*\bigr)$.

For the other implication, consider $(t_1,\ldots,t_n)$ a vector of
$\bigcap_{1\leq i\leq k} \Pi_A\bigl(h^{-1}(L_i\cap B)\cap a_1^*\cdots a_n^*\bigr)$ and let
$w=w_1^{t_1} \ldots w_n^{t_n}$. For every $1\leq i\leq k$, we will show that
$w\in L_i\cap B$. As $(t_1,\ldots,t_n)\in \Pi_A\bigl(h^{-1}(L_i\cap B)\cap
a_1^*\cdots a_n^*\bigr)$, there exists a word $w'\in a_1^*\cdots a_n^*$ such
that $\Pi_A(w')=(t_1,\ldots,t_n)$ and $h(w')\in L_i\cap B$. We conclude from
$\Pi_A(w')=(t_1,\ldots,t_n)$, that $w'=a_1^{t_1} \ldots a_n^{t_n}$ and finally
that, $h(w')=w$ belongs to $L_i\cap B$.

The class of CFLs is effectively closed under inverse
homomorphism and intersection with a regular language \cite{HU79}.  Moreover,
given a CFL, we can compute its Parikh image which is a
semilinear set. Finally, we can compute the semilinear sets
$\Pi_A\bigl(h^{-1}(L_i\cap B)\cap a_1^*\cdots a_n^*\bigr)$ and the emptiness of
the intersection of semilinear sets is decidable \cite{ginsburg}.
\qed
\end{proof}
While Lem.~\ref{lem:capcfleb} shows decidability for every elementary bounded
language, in practice, we want to select $B$ ``as large as possible''.
We select $B$ using Th.~\ref{theo-main}.
We first compute for each language $L_i$ the elementary bounded language
$B_i={w_1^{(i)}}^*\cdots{w_{n_i}^{(i)}}^*$ such that $\Pi(L_i\cap B_i)=\Pi(L_i)$. Finally, we choose $B=B_1 \cdots B_k$.

By repeatedly selecting and removing a bounded language $B$ from each $L_i$
where $1\leq i\leq k$ we obtain a sequence $\{L_i^j\}_{j\geq 0}$ of languages
such that $L_i = L_i^0 \supseteq L_i^1\supseteq \dots$.
The result of Lem.~\ref{lem:progress} shows that for each word $w\in L_i$,
there is some $j$ such that $w\notin L_{i}^{j}$, hence that 
the above sequence is strictly decreasing, that is $L_i = L_i^0 \supsetneq
L_i^1\supsetneq \dots$, and finally that if $\bigcap_{1\leq i\leq k}
L_i\neq\emptyset$ then the iteration is guaranteed to terminate.

\iflong
At Alg.~\ref{alg:cap}, we present a pseudocode for the special case of the
intersection of two CFLs.

\begin{algorithm}[t]
	\caption{Intersection\label{alg:cap}}
  \SetKwFor{Repeat}{repeat forever}{}{end}
  \KwIn{$L_1^0$, $L_2^0$ : CFLs}
  $L_1\leftarrow L_1^0$, $L_2\leftarrow L_2^0$\;
  \Repeat{}{
     \eIf{$\Pi(L_1)\cap \Pi(L_2)=\emptyset$}{\KwRet{$L_1^0\cap L_2^0$ is empty}}
     {Compute $B_1$ and $B_2$ elementary bounded languages such that $\Pi(L_1\cap B_1)=\Pi(L_1)$
     and $\Pi(L_2\cap B_2)=\Pi(L_2)$\;
     Compute $B=B_1 \cdot B_2$\;
     \If{$L_1\cap^{(B)} L_2\neq \emptyset$}{\KwRet{$L_1^0\cap L_2^0$ is not empty}}
     }
     $L_1\leftarrow L_1\cap \overline{B}$, $L_2\leftarrow L_2\cap \overline{B}$
  }
\end{algorithm}

\fi

\smallskip
\noindent{\bf Comparison with Context-Bounded Reachability.}
A well-studied under-approximation for multithreaded reachability is given by
context-bounded reachability \cite{QR}.
Given a pushdown network, global configurations $C_0$ and $C$, and a number
$k\geq 1$, the \emph{context-bounded reachability problem}
asks whether $C_0\leadsto_{k} C$ holds, i.e. if $C$ can be reached from $C_0$ in $k$ context switches.
This problem is decidable \cite{QR}.
Context-bounded reachability has been successfully used in practice for bug finding.
We show that underapproximations using bounded languages (Lem.~\ref{lem:capcfleb})
subsumes the technique of context-bounded reachability in the following sense.

\begin{proposition}
Let $\mathcal{N}$ be a pushdown network, $C_0,C$ global configurations
of $\mathcal{N}$, and $(L_1,\ldots,L_n)$ CFLs over alphabet $\Sigma$ such that
$C_0\rightarrow^* C$ iff $\cap_i L_i \not = \emptyset$.  For each $k\geq 1$,
there is an elementary bounded language $B_k$ such that $C_0 \leadsto_k C$ only
if $\bigcap_i^{(B_k)} L_i \not = \emptyset$.  Also, $\bigcap_i^{(B_k)} L_i \not =
\emptyset$ only if $C_0 \rightarrow^* C$.
\end{proposition}
\begin{proof}
Consider all sequences $C_0\leadsto C_1 \cdots C_{k-1}\leadsto C_k$
of $k$ or fewer switches.
By the CFL encoding (Prop.~\ref{prop:cflencoding}) each of these
sequences corresponds to a word in $\Sigma^k$.
If $C_0\leadsto_k C$, then there is a word $w\in \bigcap_i L_i$ and $w\in \Sigma^k$.
Define $B_k$ to be $w_1^*\cdots w_m^*$ where $w_1,\dots,w_m$ is an enumeration
of all strings in $\Sigma^k$.
We conclude from $w\in \Sigma^k$ and the definition of $B_k$ that $w\in B_k$,
hence that $\bigcap_i^{(B_k)} L_i \not = \emptyset$ since $w\in \bigcap_i L_i$.
For the other direction we conclude from
$\bigcap_i^{(B_k)} L_i \not = \emptyset$ that $\bigcap_i L_i \not = \emptyset$,
hence that $C_0\rightarrow^* C$.
\qed
\end{proof}

However, underapproximation using bounded languages
can be more powerful than context-bounded reachability in the following sense.  There is a family
$\set{(\mathcal{N}_k, C_{0k},C_k)}_{k\in\nats}$ of pushdown network reachability problems
such that $C_{0k}\leadsto_{k} C_k$ but $C_{0k} \not\leadsto_{k-1} C_k$ for each
$k$, but there is a single elementary bounded $B$ such that $\bigcap_i^{(B)}
L_{ik} \neq \emptyset$ for each $k$, where again $(L_{1k},\ldots,L_{nk})$ are
CFLs such that $C_{0k}\leadsto C_k$ iff $\cap_i L_{ik} \neq\emptyset$
(as in Prop.~\ref{prop:cflencoding}).

For clarity, we describe the family of pushdown networks as a family of
two-threaded programs whose code is shown in Fig.~\ref{fig-pushfamily}. 
The programs in the family differs from each other by the value to which $\mathtt{k}$ is instantiated: $\mathtt{k}=0,1,\dots$.
Each program has two threads.
Thread one maintains a local counter $\mathtt{c}$
starting at $\mathtt{0}$.  Before each increment to $\mathtt{c}$, thread one
sets a global $\mathtt{bit}$.  Thread two resets $\mathtt{bit}$.  The target
configuration $C_k$ is given by the exit point of $\mathtt{p1}$.  We conclude
from the program code that hitting the exit point of $\mathtt{p1}$ requires
$\mathtt{c}\geq \mathtt{k}$ to hold. For every
instance, $C_k$ is reachable, but it requires at least $k$ context
switches.  Thus, there is no fixed context bound that is sufficient
to check reachability for every instance in the family.
In contrast, the elementary bounded language given by
$\bigl((\mathtt{bit==true},2)\cdot(\mathtt{bit==false},1)\bigr)^*$ is
sufficient to show reachability of the target for {\bf every} instance in the
family.
\begin{figure}[h]
\centering
\begin{minipage}[t]{.5\linewidth}
\begin{verbatim}
thread p1() {
  int c=0;
L:bit=true;
  if bit == false { ++c; }
  if c<k { goto L; }
}
\end{verbatim}
\end{minipage}\hspace{1cm}%
\begin{minipage}[t]{.3\linewidth}
\begin{verbatim}
thread p2() {
L1:bit = false;
   goto L1;
}
\end{verbatim}
\end{minipage}
\caption{The family of pushdown network with global $\mathtt{bit}$.\label{fig-pushfamily}}
\end{figure}

\iflong
\subsection{Recursive Counter Machines}

In verification, counting is a powerful abstraction mechanism. Often,
counting abstractions are used to show decidability of the verification
problem. Counting abstractions have been applied on a wide range of
applications from parametrized systems specified as concurrent {\sc java}
programs to cache coherence protocols (see \cite{Vanb03}) and to programs
manipulating complex data structures like lists (see for instance
\cite{BB+06}). In those works, counting not only implies decidability, it
also yields precise abstractions of the underlying verification problem.
However, in those works recursion (or equivalently the call stack) is not part
of the model. One option is to abstract the stack using additional counters,
hence abstracting away the stack discipline. Because counting abstractions
for the stack yields too much imprecision, we prefer to use a precise model
of the call stack and perform an underapproximating analysis. This is what
is defined below for a model of recursive programs that manipulate
counters.

\noindent
{\bf Counter Machine: Syntax and Semantics.}
An \emph{$n$-dimensional counter machine} $M=(Q,T,\alpha,\beta,\set{G_t}_{t\in
T})$ consists of the finite non-empty sets $Q$ and $T$ of locations and
transitions, respectively;
two mappings $\alpha\colon T\mapsto Q$ and $\beta\colon T\mapsto Q$,
and a family $\set{G_t}_{t\in T}$ of semilinear (or \emph{Presburger definable}) sets over $\mathbb{N}^{2n}$.

A \emph{$M$-configuration} $(q,x)$ consists of a location $q\in Q$ and a vector $x\in\nats^n$; we define $C_M$ as the set of $M$-configurations.
For each transition $t\in T$, its semantics is given by the reachability relation $R_M(t)$ over $C_M$
defined as
$(q,x) R_M(t) (q',x')$ if{}f $q=\alpha(t)$, $q'=\beta(t)$, and $(x,x')\in G_t$.
The reachability relation is naturally
extended to words of $T^*$ by defining $R_M(\varepsilon) = \set{((q,x),(q,x))\mid (q,x)\in C_M}$
and $R_M(u\cdot v) = R_M(u) \comp R_M(v)$.
Also, it extends to languages as expected.
Finally, we write $(M,D)$ for a counter machine $M$
with an initial set $D\subseteq C_M$ of configurations.
Note that semilinear sets carry over subsets of $C_M$
using a bijection from $Q$ to $\set{1,\ldots,|Q|}$.

\smallskip %
\noindent %
{\bf Computing the Reachable Configurations.}
Let $R\subseteq C_M\times C_M$ and $D\subseteq C_M$, we
define the set of configurations $\post[R](D)$ as
$\set{(q,x)\mid \exists (q_0,x_0)\in D \land (q_0,x_0)R(q,x)}$. Given a $n$-dim
counter machine $M=(Q,T,\alpha,\beta,\set{G_t}_{t\in T})$, a semilinear set $D$
of configurations and a CFL $L\subseteq T^*$
(encoding execution paths), we want to underapproximate $\post[R_M(L)](D)$:
the set of $M$-configurations reachable from $D$ along words of $L$.
Our underapproximation computes the set $\post[R_M(L')](D)$ where $L'$ is a
Parikh-equivalent bounded subset $L$ such that $L'=L\cap B$ where
$B=w_1^*\cdots w_n^*$.

We will construct, given $(M, D)$, $L$ and $B$ (we showed above how to
effectively compute such a $B$), a pair $(M',D')$ such that the set of
$M$-configurations reachable from $D$ along words of $L\cap B$ can be
constructed from the set of ${M'}$-configurations reachable from $D'$.  Without
loss of generality, we assume $M$ is such that $Q$ is a singleton.  (One can
encode locations using counters.)

Let $M=(Q,T,\alpha,\beta,\set{G_t}_{t\in T})$ a $\gamma$-dim counter machine with
$Q=\set{q_f}$ and $B=w_1^*\cdots w_n^*$ such that $\Pi(L\cap B)=\Pi(L)$. Let
$h$ be the homomorphism that maps some fresh symbols $a_1,\ldots,a_n$ to the
words $w_1,\ldots,w_n$, respectively. We compute the language
$L'_A=h^{-1}(L\cap B)\cap a_1^*\cdots a_n^*$. Let
$S=\Pi_{\set{a_1,\ldots,a_n}}(L'_A)$, and note that $S$ is a semilinear set.
For clarity, we first consider a linear set $H$ where
$p_0=(p_{01},\ldots,p_{0n})$ denotes the constant and
$\set{p_i=(p_{i1},\ldots,p_{in})}_{i\in I\setminus \set{0}}$ the set of periods
of $H$ and $I=\set{0,\ldots,k}$.  Let $J=\set{1,\ldots,n}$.  In the following,
for every pair of vectors $x=(x_1,\ldots,x_r)$ and $y=(y_1,\ldots,y_s)$, we
denote by $(x,y)$ the vector $(x_1,\ldots,x_r,y_1,\ldots,y_s)$.
The machine $M'$ is defined in Fig.~\ref{fig:cntauto}.
\begin{figure}[t]
	\begin{minipage}[t]{.50\textwidth}
\textbullet{} $\gamma'=\gamma+(k+1)n$\\
\textbullet{} $Q'=\set{q_i}_{0\leq i\leq k}\cup \set{q_{ij}}_{0\leq i\leq k}^{1\leq j\leq n}\cup \set{q_f}$\\
\textbullet{} $T'=\set{t_0}\cup \set{t_i,t_i^s}_{1\leq i\leq k}\cup\set{t_{ij},t_{ij}^s}_{0\leq i\leq k}^{1\leq j\leq n}$\\
\textbullet{} $\alpha'$ and $\beta'$ are given by the automaton\\
\textbullet{} Let $i\in\set{0,\dots,k}$ and $j\in\set{1,\dots,n}$\\
$G_{t_i}=\begin{cases}
	G^{+}_{\lambda_{01}} \comp \ldots \comp G^{+}_{\lambda_{0n}}&\text{if $i=0$}\\
	\{(x,x)\in\nats^{2\gamma'}\}&\text{else}
	\end{cases}$\\
$G_{t_i^s}= G^{+}_{\lambda_{i1}} \comp \ldots \comp G^{+}_{\lambda_{in}}$,\\ 
$G_{t_{ij}^s}=G(w_j^{p_{ij}})\comp G^{-}_{\lambda_{ij}}$, and\\
$G_{t_{ij}}=\{\bigl((x,v),(x,v)\bigr)\mid v_{i*n+j}=0\}$\\
\end{minipage}\hspace{.5em}%
\begin{minipage}[t]{.5\textwidth}
	\vspace{0pt}
    \includegraphics[width=\textwidth]{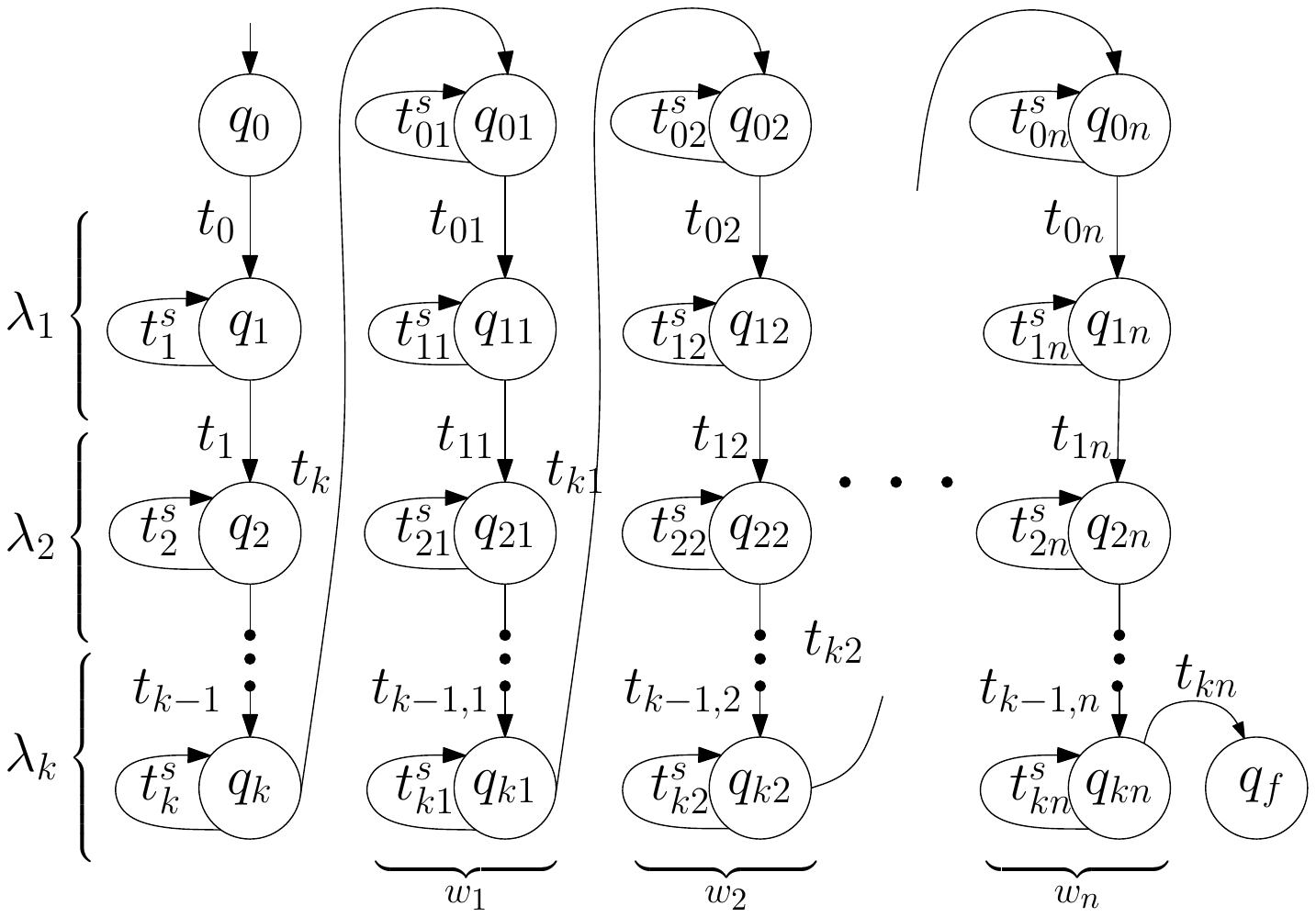}
\end{minipage}
Let $\#\in\set{+,-}$, $G^{\#}_{\lambda_{ij}}=\{\bigl((x,v),(x,v')\bigr)\in\nats^{2\gamma'}\mid v'=v\#\punit_{i*n+j}\}$.\\
Let $w\in T^*$,
$G(w)$ is s.t.
$G(\varepsilon)=\{(x,x)\in\nats^{2\gamma'}\}$,
$G(t)=\{\bigl((x,v),(x',v)\bigr)\in\nats^{2\gamma'}\mid (x,x')\in G_t\}$,
and $G(w_p \cdot w_s)=G(w_p)\comp G(w_s)$ if $w=\varepsilon$, $t$ and $w_s\cdot w_p$, respectively.
  \caption{The $\gamma'$-dim counter machine $M'=(Q',T',\alpha',\beta',\set{G_t}_{t\in T'})$.\label{fig:cntauto}}
\end{figure}

Between $q_0$ and $q_{01}$, $M'$ non-deterministically picks values for all the
additional counters which we denote $\set{\lambda_{ij}}_{i\in I,j\in J}$.
When $M'$ fires $t_k$,
we have for all $i\in I$ and $j,j'\in J$: $\lambda_{ij}=\lambda_{ij'}$ and
$\lambda_{0i}=1$.
Below, for every $i\in I$, we denote by $\lambda_i$ the common value of the
counters $\set{\lambda_{ij}}_{j\in J}$.
Then, $M'$ simulates the behavior of $M$ for the sequence of transitions
given by
$w_1^{p_{01}+\lambda_{1}p_{11}+\cdots+\lambda_{k}p_{k1}} \ldots w_n^{p_{0n}+\lambda_{1}p_{1n}+\cdots+\lambda_{k}p_{kn}}$
the Parikh image of which is $p_0+\sum_{i\in I}\lambda_i p_i$.
Let us define the set $D'$ of configurations of $C_{M'}$ as
$
	\set{(q_0,(x,v))\mid (q_f,x)\in D\land v=0^{(k+1)n}}
$.

A sufficient condition for the set of reachable configurations  of $M'$
starting from $D'$ to be effectively computable is that for each $t$ in
$\set{t_i^s}_{i\in I\setminus\set{0}}\cup\set{t_{ij}^s}_{i\in I,j\in J}$ (i.e.
the loops in Fig.~\ref{fig:cntauto}), it holds that $t^*$ is computable and
Presburger
definable. Given $t$ the problem of deciding if $t^*$ is Presburger definable
is undecidable \cite{BFLS05}.  However, there exist some subclasses $C$ of Presburger
definable sets such that if $t\in C$ then $t^*$ is Presburger definable and
effectively computable, hence the set of reachable configurations of $(M',D')$
can be computed by quantifier elimination in Presburger arithmetic. A known
subclass is that of guarded command Presburger relations.  An $n$-dimensional \emph{guarded command} 
is given by the closure under composition of
$\set{(x,x')\in \nats^{2n} \mid x'=x+\punit_i}$ (increment), $\set{(x,x')\in\nats^{2n}\mid x'=x-\punit_i}$ (decrement) and 
$\set{(x,x)\in\nats^{2n}\mid
x=(x_1,\dots,x_n) \land x_i=0}$ ($0$-test) for $1\leq i \leq n$.


Other subclasses are
given in \cite{BozgaGI09,FL02}.
Note that if for each $t\in T$ of $M$, $G_t$ is given by a guarded command
then so is each $G_{t'}$ for $t'\in T'$ of $M'$ by definition.

Hence, we find that the set $\post[R_{M'}(T'^{*})](D')$ of reachable
configurations of $(M',D')$ is Presburger definable, effectively computable and
relates to $\post[R_M(L')](D)$ for the bounded language $L'$ as follows.
%
\begin{lemma}
	Let $(q_f,x)\in C_{M}$,\\
	$(q_f,x)\in \post[R_M(L')](D)$ if{}f 
	$\exists v\in\nats^{(k+1)n}\colon (q_f,(x,v))\in \post[R_{M'}(T'^{*})](D')$.
\end{lemma}
We can easily compute the intersection of the two semilinear sets $S$ and $\set{q_f}\times\nats^{\gamma}$ over $Q'\times \nats^{\gamma}$, because of the way we have carried the notion of semilinear set over $Q'\times \nats^{\gamma}$. We take a bijection $\eta$ from $Q'$ to $\set{1,\ldots,|Q|}$, so a configuration $(q,x)\in Q'\times \nats^{\gamma}$ is represented by $(p_1,\ldots,p_{|Q|},x)^T$ with $p_j=\begin{cases} 1 &\text{if } \eta(q)=j\\ 0 &\text{ otherwise} \end{cases}$. Hence, the intersection consists of all the vectors of $S$ with the composant of $q_f$ equal to one and the others equal to zero.
Lem.~\ref{lem:progress} shows that by iterating the construction we obtain a
semi-algorithm for a context-free language.
\fi

\noindent
{\bf Acknowledgment.}
We thank Ahmed Bouajjani for pointing that the bounded languages approach
subsumes the context-bounded switches one.

\end{document}